\documentclass[final,5p,times,twocolumn,authoryear]{elsarticle}

\usepackage{flushend}
\usepackage{graphicx}
\usepackage[authoryear,round]{natbib}
\usepackage{tikz}
\usepackage{float}
\usepackage{fancyhdr}

\usepackage{nomencl}    
\usepackage{multicol}   
\usepackage{framed}     
\usepackage{titlesec}    
\usepackage{enumitem}    

\makenomenclature

\renewcommand*\nompreamble{\begin{multicols}{2}} 
	\renewcommand*\nompostamble{\end{multicols}}     

\usepackage{amssymb}
\usepackage{amsmath}

\DeclareRobustCommand\sampleline[1]{%
	\tikz\draw[#1,very thick] (0,0) (0,\the\dimexpr\fontdimen22\textfont2\relax)
	-- (1em,\the\dimexpr\fontdimen22\textfont2\relax);%
}

\definecolor{colmm1}{rgb}{0.0000, 0.4470, 0.7410}
\definecolor{colmm2}{rgb}{0.8500, 0.3250, 0.0980}
\definecolor{colmm3}{rgb}{0.9290, 0.6940, 0.1250}
\definecolor{colmm4}{rgb}{0.4940, 0.1840, 0.5560}
\definecolor{colmm5}{rgb}{0.4660, 0.6740, 0.1880}
\definecolor{colmm6}{rgb}{0.3010, 0.7450, 0.9330}
\definecolor{colmm7}{rgb}{0.6350, 0.0780, 0.1840}

\journal{}

\begin{document}
	
	\begin{frontmatter}
		
		
		
\title{The three effects of the pressure force on turbulent boundary layers} 
		
		\author[label1,label2]{Taygun Recep Gungor}
		\author[label2]{Ayse Gul Gungor} 
		
		\affiliation[label1]{organization={Department of Mechanical Engineering},
			addressline={Universite Laval},
			city={Quebec City},
			postcode={G1V 0A6},
			state={QC},
			country={Canada}}
		
		\affiliation[label2]{organization={Faculty of Aeronautics and Astronautics},
			addressline={Istanbul Technical University},
			city={Istanbul},
			postcode={34469},
			country={Turkey}}
		
		\author[label1]{Yvan Maciel} 

		\begin{abstract}
This study aims to isolate the three effects of the pressure force on the inner and outer layers: the local direct impact (characterized by the pressure gradient parameter, $\beta$), the local disequilibrating effect (represented here by the normalized streamwise derivative $d\beta/dX$), and the upstream cumulative effect, while also accounting for the inevitable Reynolds number influence. \color{black} To achieve this objective, we draw on several non-equilibrium and near-equilibrium databases from the literature, and employ a methodology based on the selection of pressure-gradient parameters—distinct for the inner and outer layers—that capture the local direct impact and the local disequilibration effect of the pressure gradient. \color{black} The pressure force impact on the inner and outer regions is represented by two parameters: the friction-viscous pressure-gradient parameter, $\beta_i$, and the pressure-gradient parameter based on Zagarola-Smits velocity, $\beta_{ZS}$, respectively. In the non-equilibrium flow cases, both $\beta_i$ and $\beta_{ZS}$ exhibit similar distributions, initially increasing and then decreasing. However, the rate of change of these parameters along the streamwise direction varies among the flows, indicating differing levels of pressure force disequilibration. In addition, we employ two near-equilibrium cases with minimal variations of the pressure gradient parameters for comparisons. In the outer layer, it is found that both the local and cumulative disequilibrating effects modify the mean velocity and Reynolds stress profiles at identical $\beta_{ZS}$ values. The faster the variations in pressure force impact, the more delayed the response of both mean flow and turbulence. Cumulative effects prove to be significant. In the inner layer, which responds much faster to changes in pressure force, the local disequilibrating effect still modifies the mean velocity profile in the viscous sublayer. Notably, when the mean velocity defect is significant, the behavior of $u$-structures in the inner layer appears to be governed by how outer turbulence responds to pressure force effects. In contrast, the size of $uv$-structures in the inner layer scales with the mixed pressure-friction length $(\nu/u_{\tau}+\nu/u^{pi})$. Unlike inner $u$-structures, they are independent of large-scale outer structures and flow history.
		\end{abstract}
		
		\begin{graphicalabstract}
		\end{graphicalabstract}
		
		\begin{highlights}
			
			\item Pressure force variations alter the flow even when the force balance is similar.
			\item Rapid pressure force variations further delay the flow response.
\item Flow history has minimal effect on inner velocity, regardless of pressure force.
\item Inner-layer $uv$-structures size scales with a mixed friction-pressure length scale
		\end{highlights}
		
		
	\end{frontmatter}


\nomenclature[A]{$Re$}{Reynolds number}
\nomenclature[A]{$Pr$}{Prandtl number}
\nomenclature[A]{$Nu$}{Nusselt number}
\nomenclature[G]{$\mu$}{Dynamic viscosity [Pa·s]}
\nomenclature[G]{$\rho$}{Density [kg/m\textsuperscript{3}]}
\nomenclature[S]{$i$}{Index representing inlet}
\nomenclature[S]{$w$}{Wall condition}
\nomenclature[Q]{$'$}{Time derivative}
\nomenclature[Q]{$*$}{Non-dimensional quantity}


\onecolumn

\section*{Nomenclature}
\vspace{-1em}
\begin{framed}
	\begin{multicols}{2}
		
		\subsection*{Roman Symbols}
		\begin{description}[leftmargin=1.5em, style=multiline, labelwidth=1.2em]
\item[$x$]  The streamwise coordinate
\item[$y$]  The wall-normal coordinate
\item[$z$]  The spanwise coordinate

\item[$u$]  Velocity fluctuation in the streamwise direction
\item[$v$]  Velocity fluctuation in the wall-normal direction
\item[$w$]  Velocity fluctuation in the spanwise direction
\item[$U$]  Mean velocity in the streamwise direction
\item[$V$]  Mean velocity in the wall-normal direction

			\item[$f$] Frequency
			\item[$k$] Wavenumber
			\item[$p$] Pressure
			\item[$u_\tau$] Friction velocity
		\end{description}
		
		\subsection*{Greek Symbols}
		\begin{description}[leftmargin=1.5em, style=multiline, labelwidth=1.2em]
\item[$\nu$] Viscosity 
\item[$\beta$] Pressure gradient parameter
\item[$\beta_C$] Clauser pressure gradient parameter
\item[$\beta_{ZS}$] Pressure gradient parameter based on Zagarola-Smits scales	
\item[$\beta_{i}$] Inner layer pressure gradient parameter 
\item[$\rho$] Density
\item[$\delta$] Boundary layer thickness
\item[$\delta^*$] Displacement thickness
\item[$\theta$] Momentum thickness

		\end{description}
		
		\subsection*{Subscripts}
		\begin{description}[leftmargin=1.5em, style=multiline, labelwidth=1.2em]
			\item[$e$] Boundary layer edge
			\item[$w$] Wall
			\item[$ZS$] Zagarola-Smits scaling
			\item[$av$] Averaged over the whole domain
			\item[$i$] Inner layer
		\end{description}
		
		\subsection*{Superscripts}
		\begin{description}[leftmargin=1.5em, style=multiline, labelwidth=1.2em]
			\item[$+$] Friction-viscous scale
			\item[$pi$] Pressure-viscous scale
			\item[$*$] Mixed scale
		\end{description}
		
	\end{multicols}
\end{framed}

	\twocolumn

				\color{black} 
		\section{Introduction}
		\label{sec1}

The development of a turbulent boundary layer (TBL) under the influence of an adverse pressure gradient (APG) exhibits distinct characteristics compared to those observed in canonical wall-bounded flows, such as channel and pipe flows, or TBLs with zero pressure gradient. This difference becomes particularly pronounced under conditions of a strong pressure gradient or prolonged exposure to it, significantly altering the mean flow behavior and, subsequently, the turbulence characteristics. The technological relevance of APG TBLs has spurred numerous studies aimed at elucidating the impact of APG on	flow dynamics and turbulence. A favorable pressure gradient (FPG) generally has a less significant impact on the turbulent boundary layer. Consequently, FPG TBLs have received less attention than APG TBLs, though they remain relevant. While this study primarily examines the effects of the APG, we also consider one FPG case.

		
The major differences between APG TBLs and canonical wall-bounded flows have been well-reported in the literature. As the APG acts on the flow, the mean shear increases in the outer layer and decreases in the inner layer. Consequently, turbulence in the outer layer increases compared to that in canonical flows, \color{black} when normalized with friction velocity or edge velocity\color{black}. In large-defect APG TBLs, outer-layer turbulence becomes dominant \citep{skaare1994turbulent,gungor2016scaling}. Outer-layer turbulence also increases in canonical wall flows with Reynolds number, but its properties and generating mechanisms differ \citep{vila2020separating}. Contrary to what happens in the outer layer, inner-layer turbulence becomes weaker with increasing momentum deficit.

In APG TBLs with a small velocity defect, inner-layer turbulence remains very similar to that in canonical flows \citep{gungor2024turbulent}; however, the differences become pronounced in large-defect TBLs. If the APG is sufficiently strong or prolonged, the inner peak of the $\langle u^2\rangle$ spectra ---characteristic of inner-layer streaks--- disappears \citep{kitsios2017direct,lee2017large,gungor2022energy}. 
		
Despite growing efforts to understand the nature of pressure gradient (PG) TBLs, many questions about the effects of the pressure force remain unanswered. \cite{gungor2024turbulent} discuss how the pressure force affects the TBL in three distinct ways: 
\begin{enumerate}
	\item Local direct impact: the immediate action of the pressure force on the fluid, reflected in the force balance.
	\item Local disequilibrating effect: a local change in the pressure force that alters the force balance.
	\item Upstream history effect: the contribution of the upstream pressure force distribution to the local state of the boundary layer.
\end{enumerate}

\color{black}

These three effects are closely linked to the concept of equilibrium in turbulent boundary layers. In the context of pressure-gradient TBLs, equilibrium implies that the balance of forces acting on the fluid remains unchanged as the flow develops downstream \citep{rotta1953theory,clauser1956turbulent,maciel2006self}. It is important to emphasize that, in an equilibrium TBL, flow properties such as boundary layer thickness, velocity, and force magnitudes may vary in the streamwise direction; however, the ratios between the governing forces remain constant. The flow is thus in a state of dynamic equilibrium, or similarity. For a more detailed discussion on equilibrium conditions in TBLs, see \cite{maciel2006self} and \cite{devenport2022equilibrium}.

With respect to the three pressure force effects discussed earlier, an equilibrium TBL is characterized by the presence of only the local direct impact, with both the local disequilibrating effect and the upstream cumulative (history) effect being absent. Except for the idealized sink-flow TBL \citep{townsend1976structure,rotta1962turbulent}, complete equilibrium of TBLs is not achievable at finite Reynolds numbers due to the coexistence of two dynamically distinct wall-normal layers—the inner and outer layers—each governed by its own characteristic length scale. In practice, both experiments and simulations can only approximate equilibrium conditions, generating what are referred to as \emph{near-equilibrium} TBLs. In such flows, the force balance varies slowly in the streamwise direction, due to the gradual increase in the local Reynolds number. A canonical example is the zero-pressure-gradient (ZPG) TBL, which exhibits near-equilibrium behavior at finite Reynolds numbers. In the present study, datasets of near-equilibrium TBLs are used to isolate the local direct impact from the other pressure force effects.

\color{black}


A few studies have investigated history effects using flows with varying flow histories. \cite{vinuesa2017revisiting} examined  flow history effects using various databases to assess the behaviour of APG TBLs. In a similar fashion, \cite{bobke2017history} investigated existing databases in the literature at specific streamwise positions where Clauser pressure gradient parameter, $\beta_{C}$ (equation \ref{pg1}), and the friction Reynolds number $Re_{\tau}$ match to isolate the effect of flow history to some extent.  
		
		\begin{equation}
			\beta_{C} = \frac{\delta^*}{\rho u_\tau^2}\frac{dp_e}{dx}
			\label{pg1}
		\end{equation}
		
\noindent Here $\delta^*$ is the displacement thickness, $\rho$ is density, $u_\tau$ is the friction velocity and $p_e$ is the static pressure at the edge of the boundary layer. The pressure gradient parameter $\beta_{C}$ quantifies the ratio of the streamwise pressure force to the wall friction force influencing the entire boundary layer momentum defect. As a global parameter, it does not distinguish between pressure gradient effects in the inner and outer layers, which differ significantly. For that reason, \cite{gungor2024turbulent} utilized $\beta_i$ and $\beta_{ZS}$ (equation \ref{pg2}) as the selected pressure-gradient parameters in the inner and outer layers, respectively, which reflects the local force balance in each layer \citep{maciel2018outer,gungor2024turbulent}.

		\begin{equation}
			\beta_i = \frac{v}{\rho u_\tau^3}\frac{dp_w}{dx}, \hspace{0.3cm} \beta_{ZS}=\frac{\delta}{\rho U_{ZS}^2}\frac{dp_e}{dx}.
			\label{pg2}
		\end{equation}
		
\noindent Here $\nu$ is viscosity, $p_w$ is pressure at wall and, $\delta$ is boundary layer thickness, and $U_{ZS}$ is the  Zagarola–Smits velocity scale, defined as follows

		\begin{equation}
U_{ZS}= U_e \frac{\delta^*}{\delta}
\end{equation}

\noindent where $U_e$ is the velocity at the edge of the boundary layer. 
		
While existing literature attempts to shed light on this matter, there is an evident gap in adopting a systematic approach to rigorously explore this issue. The databases in the literature often have arbitrary PG distributions with varying features. Moreover, documented flow cases typically exhibit small velocity defects, rendering the isolation of pressure gradient effects from those attributable to Reynolds number variations as particularly challenging.
		
			This study seeks to delineate the influence of flow history, as well as the local disequilibrium effect of the pressure force, on mean flow and turbulence in non-equilibrium PG TBLs reaching large velocity defects. \color{black} To this end, we analyze a set of non-equilibrium and near-equilibrium datasets from the literature, using a methodology based on the selection of pressure-gradient parameters—distinct for the inner and outer regions—that isolate the local direct influence and the local disequilibration effect of the pressure gradient. \color{black} The selected databases span a broad spectrum of velocity defect conditions from very-small defect to large defect of TBLs close to separation. One database also includes an FPG TBL with a prolonged APG upstream history. This allows us to examine the effect of flow history in different flow conditions. We study the mean flow, Reynolds stresses, and spectral distributions along with the local force balance to understand the pressure force effects.

		\begin{figure*}[h!]
			\centering 
			
			\begin{tikzpicture}
				\node(a){ \includegraphics[scale=0.75]{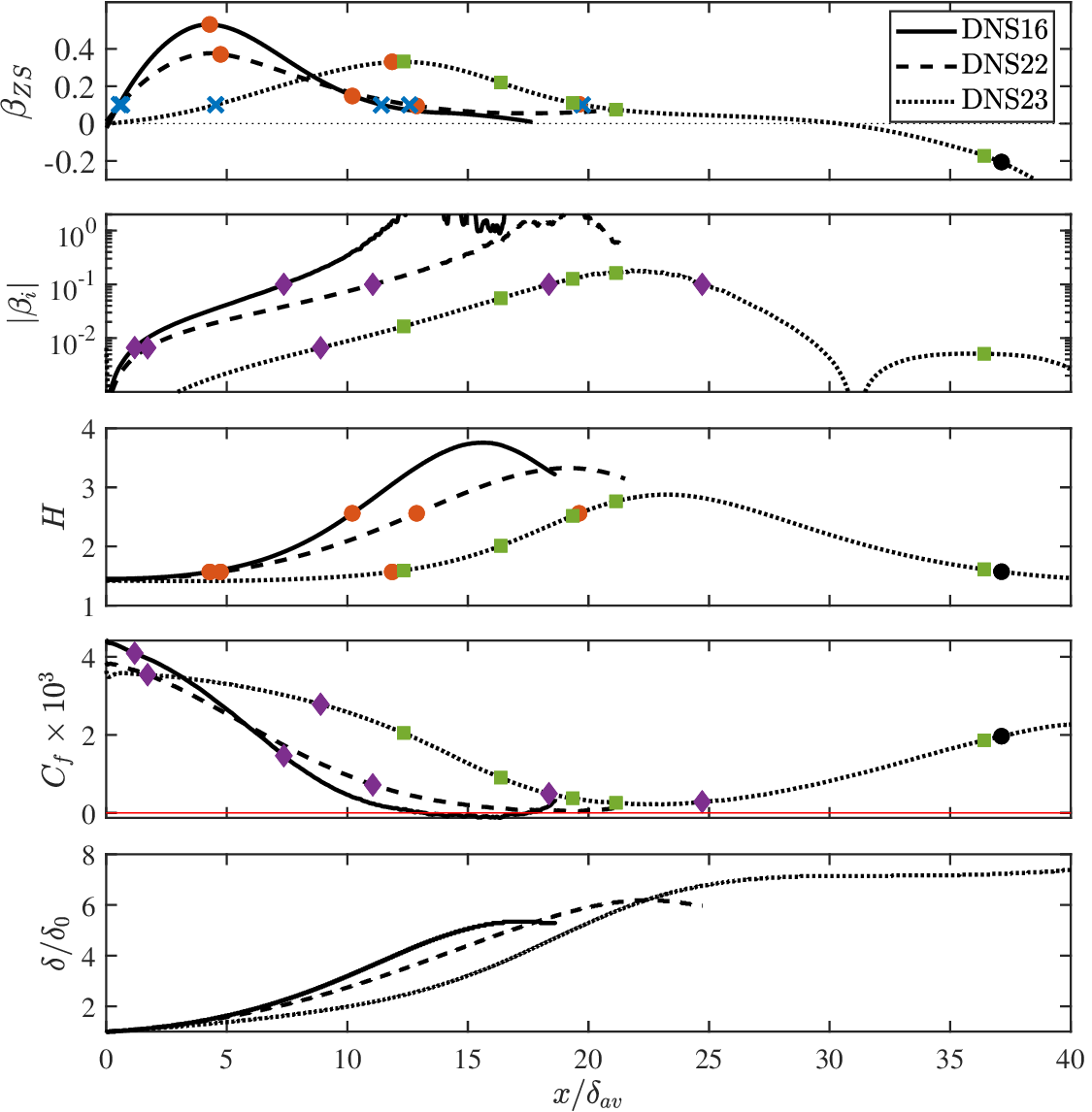}};
				\node at (-7.7,6.5) {\large $(a)$};
				\node at (-7.7,3.6) {\large $(b)$};
				\node at (-7.7,1) {\large $(c)$};
				\node at (-7.7,-2) {\large $(d)$};
				\node at (-7.7,-5) {\large $(e)$};
								\draw[black, very thick] (4.05,2.2)  -- (4.05,3.5);
				\draw[black, very thick,<->] (4.05,3.0) -- (6.8,3);
				
								\node at (5.3,3.5) {\color{black}{$\beta_i < 0$}};
				
							\end{tikzpicture}
			\caption{The $\beta_{ZS}$, $\beta_{i}$, $H$, $C_f$, and $\delta$ distributions of the non-equilibrium flow cases as a function of $x/\delta_{av}$. The markers on the lines indicate the positions examined throughout the paper, as described in the text.}
			\label{figure1}
		\end{figure*}
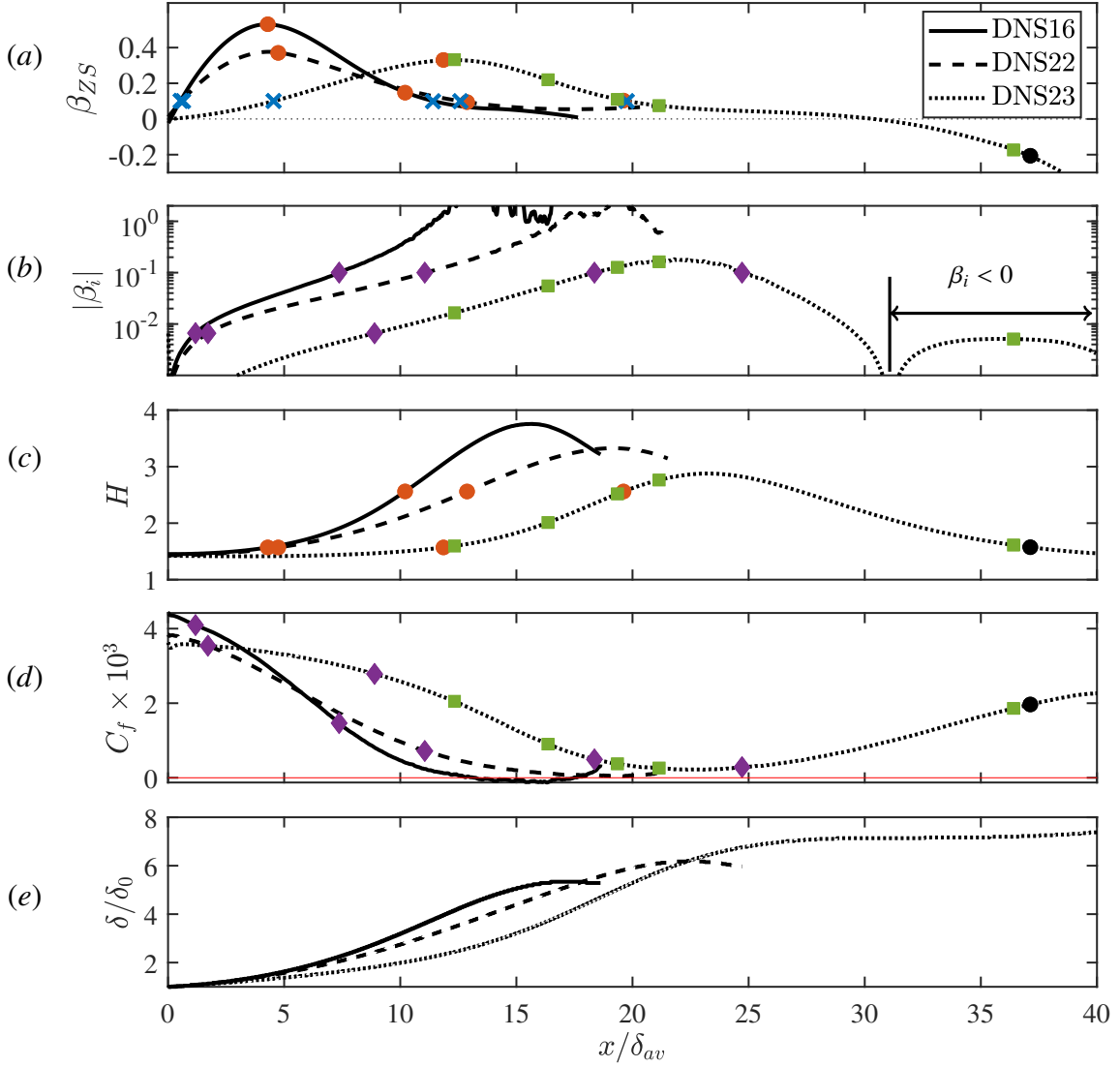

		\section{Methodology}
		\label{sec2}


		\subsection{The databases}

For this paper, we utilize five databases from the literature. The first three correspond to highly non-equilibrium flows. In addition, we choose two near-equilibrium cases to have TBLs where the local direct impact of the pressure force is dominant. Thus, in these TBLs, the local disequilibrating and upstream history effects are minimal, clarifying the distinction between local direct impact and the other effects.

		\begin{figure*}[h!]
			\centering
			\vspace{-0.5cm}

	\begin{tikzpicture}
	\node(a){ \includegraphics[scale=0.6]{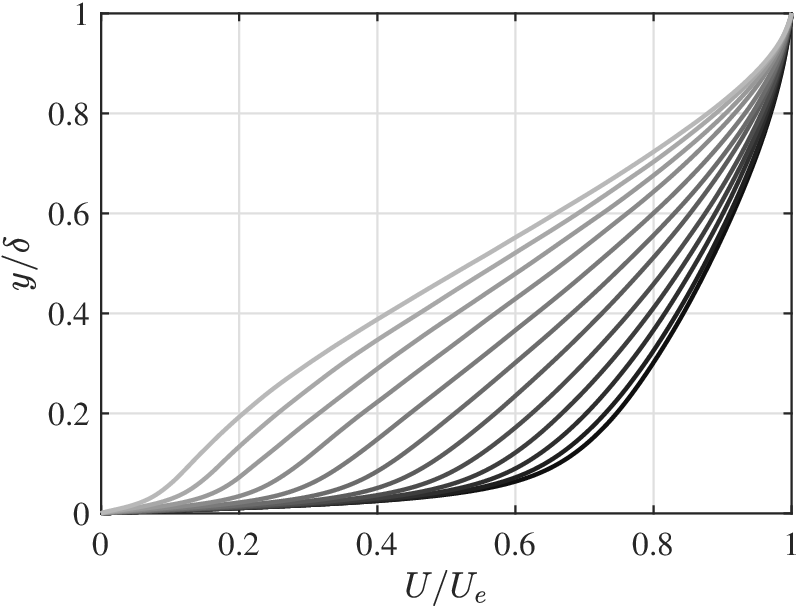}};
	\node at (-3.9,2.8) {\large $(a)$};
	\end{tikzpicture} 	\begin{tikzpicture}
	\node(a){ \includegraphics[scale=0.6]{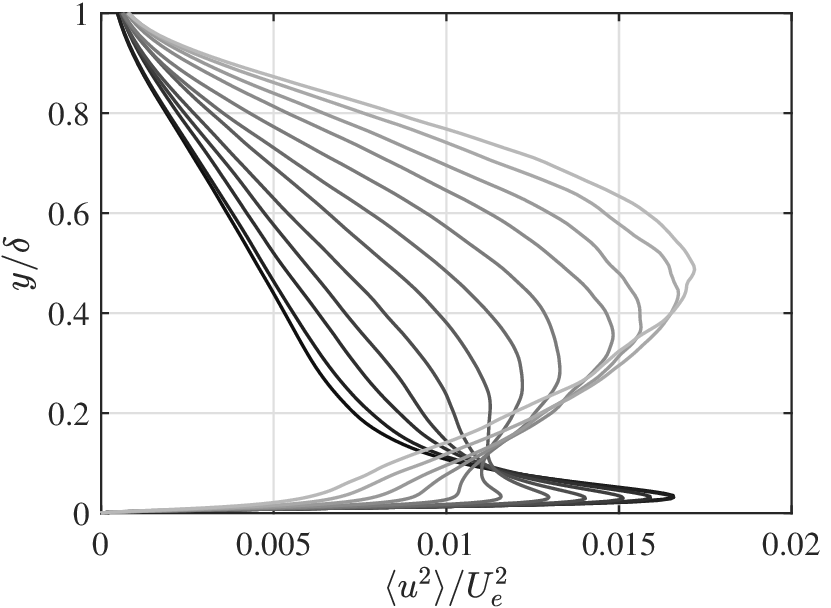}};
		\node at (-3.9,2.8) {\large $(b)$};
\end{tikzpicture}

	\begin{tikzpicture}
	\node(a){ \includegraphics[scale=0.6]{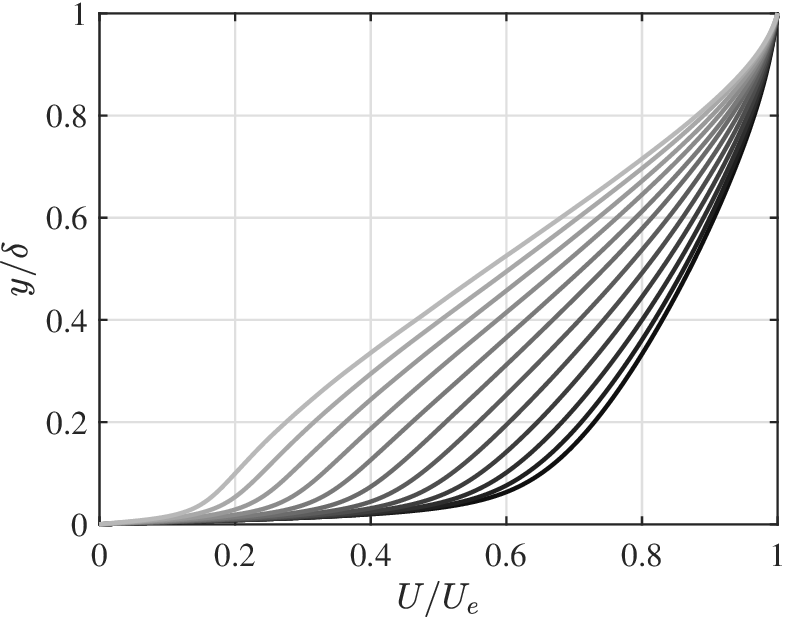}};
	\node at (-3.9,2.8) {\large $(c$};
	\end{tikzpicture} 	\begin{tikzpicture}
	\node(a){ \includegraphics[scale=0.6]{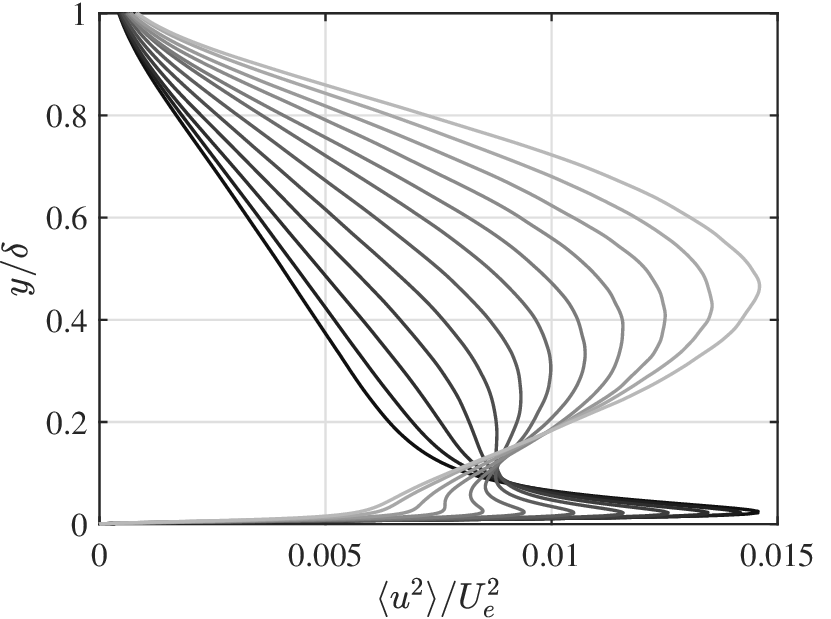}};
		\node at (-3.9,2.8) {\large $(d)$};
\end{tikzpicture}

	\begin{tikzpicture}
	\node(a){ \includegraphics[scale=0.6]{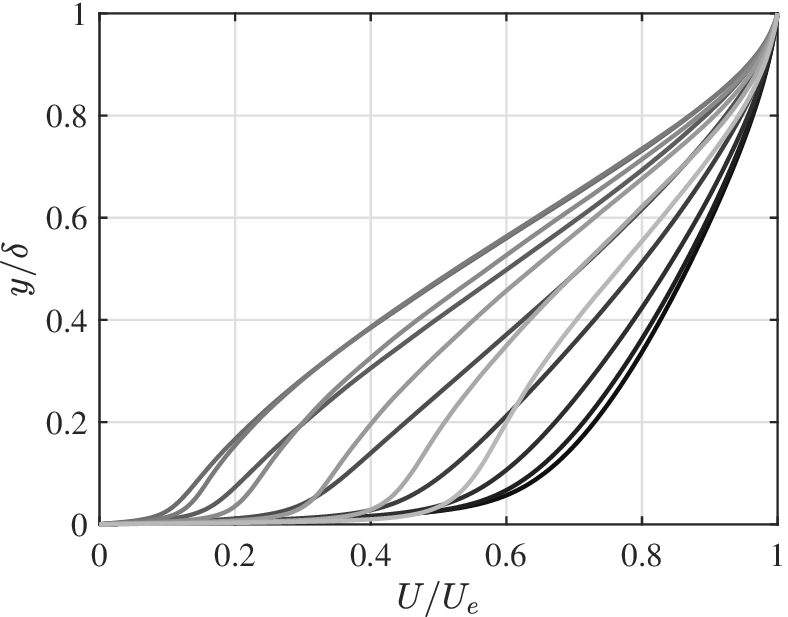}};
	\node at (-3.9,2.8) {\large $(e)$};
	\end{tikzpicture} 	\begin{tikzpicture}
	\node(a){ \includegraphics[scale=0.6]{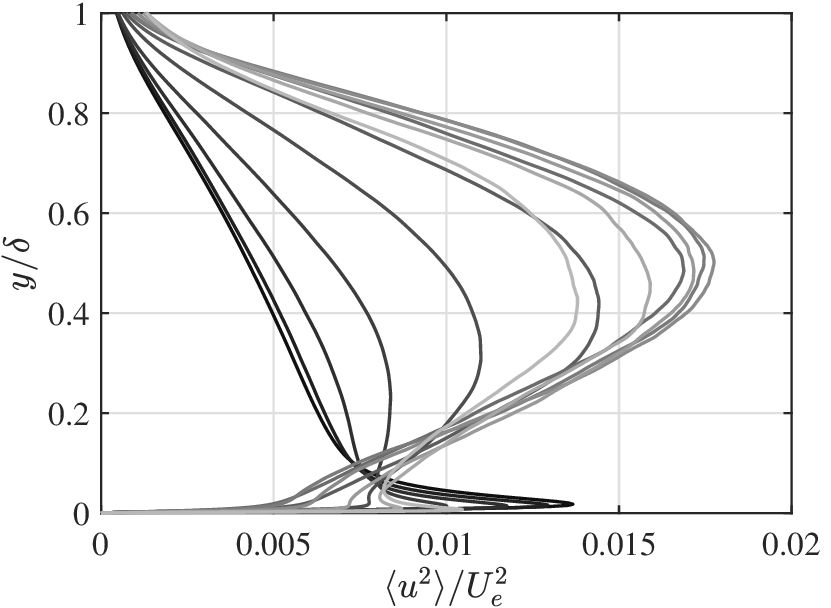}};
		\node at (-3.9,2.8) {\large $(f)$};
\end{tikzpicture}
			
			\caption{The evolution of the outer-scaled mean velocity (left) and $\langle u^2\rangle$ (right) profiles of DNS16 (top), DNS22 (middle) and DNS23 (bottom).	Profiles at equidistant streamwise positions, shown only within the region investigated in this study. The regions extent is 11.5$\delta_{av}$ for DNS16, 13$\delta_{av}$ for DNS22 and 37$\delta_{av}$ for DNS23. The flow evolves from dark to light gray.}
			
			\label{son_figure}
		\end{figure*}

The first two non-equilibrium databases (DNS16 and DNS22) are the APG TBLs of \cite{gungor2016scaling} and  \cite{gungor2022energy}. The Reynolds number based on the momentum thickness ($Re_\theta$) of these cases reaches 4650 and 8650, respectively. The third one (DNS23) is the flow case of \cite{gungor2024turbulent}. Differently from the previous two TBLs, DNS23 is exposed to an APG first, followed by a FPG. The Reynolds number for this flow is the highest, with $Re_\theta$ ranging from $1940$ to $13000$. The first two flows, as well as the initial APG region of the third, exhibit an initial increase followed by a decrease in pressure force impact, resulting in continuous momentum loss. However, in DNS23, the flow regains momentum even slightly upstream of the FPG region. 

\color{black}


\color{black}
	
	In addition to these non-equilibrium TBLs, we also include the two near-equilibrium cases from \cite{kitsios2017direct} with mild and strong APGs (EQ1 and EQ2, respectively). These flows are not in full equilibrium, but the changes in the force balance within the inner and outer layers are minimal.
		
		\color{black}

Prior to discussing the flows, it is essential to define the boundary layer parameters. In contrast to canonical wall-bounded flows, the definition of the boundary layer thickness $\delta$ is less straightforward in non-equilibrium PG TBLs due to the spatial variations in the freestream velocity. The boundary layer thickness $\delta$ is defined following the approach of Wei and Knopp (2023) for all cases considered. Specifically, $\delta$ is taken as the wall-normal location where the Reynolds shear stress reaches $5\%$ of its maximum value, as this criterion provides the most consistent definition among the available methods. A more detailed discussion of this choice can be found in \cite{gungor2024turbulent}. The displacement thickness and momentum thickness are computed using the incompressible formulations: $\delta^* = \int_0^\delta \left(1 - \frac{U(y)}{U_e}\right) \, dy$, $\theta = \int_0^\delta \frac{U(y)}{U_e} \left(1 - \frac{U(y)}{U_e}\right) \, dy$, with $\delta$ as the upper limit of integration. The edge velocity $U_e$ and edge static pressure $p_e$ are the values of $U$ and $p$ at $y=\delta$.


Another important aspect is the definition of the inner and outer layers, which remains challenging for TBLs under strong APGs. In fact, a universally accepted wall-normal layer structure does not exist for large-defect TBLs. In this study, the term “outer layer” is used loosely to refer to the region above $y/\delta = 0.15$, which approximately marks the height beyond which viscous forces become negligible in all cases examined (see the mean momentum budgets in figure 4). This “outer layer” corresponds to what is commonly referred to as the wake layer in canonical wall-bounded flows, as it excludes the overlap layer —an integral part of both the inner and outer regions in classical boundary layer theory. Accordingly, the upper limit of the inner layer is taken as $y/\delta = 0.15$, the point at which viscous effects become negligible.

		\color{black}

		\subsection{Selection of the pressure-gradient parameters}
		
As stated previously, the flow dynamics in non-equilibrium TBLs are significantly influenced by three distinct pressure gradient effects \citep{gungor2024turbulent}: the local pressure force, its variation in the streamwise direction (local disequilibrating effect), and upstream flow history. To discuss the various pressure gradient effects, it is useful to present the streamwise mean momentum equation, where each term can be considered as a force acting on the flow:
		
		\vspace*{-0.5cm}
		
		\color{black}
		\begin{equation}
			0 = 
			\bigg (-U \frac{\partial U}{\partial x}-V\frac{\partial U}{\partial y} \bigg )-
			\frac{1}{\rho}\frac{dp_e}{dx}-
			\frac{\partial  \langle u v\rangle}{\partial y}+
			\frac{\partial  \langle u u\rangle}{\partial x
			}-		\nu\frac{\partial^2 U}{\partial y^2}
			\label{mombud}
		\end{equation}
		
		\vspace*{-0.25cm}
		
		\noindent where $U$ and $V$ are the mean velocities in the streamwise and wall-normal directions, $u$ and $v$ are the fluctuation velocities, and $\langle.\rangle$ indicates ensemble averaging. The terms on the right-hand side of the equation represent the inertia force, the pressure force, the tangential and streamwise normal components of the apparent turbulent force and the viscous force, respectively. 
				\color{black}

		The force balance in the inner and outer layers can be significantly different from each other and due to this, we examine the force balance in each layer separately. For the outer layer, we employ $\beta_{ZS}$ which accurately depicts the ratio of the pressure force to the turbulent force in the outer layer in APG TBLs \citep{maciel2018outer, gungor2024turbulent}. As for the inner layer, we utilize the inner-layer PG parameter, $\beta_i$, also denoted $\Delta p^+$ in the literature, which represents the ratio of pressure force to turbulent force within the inner layer. The values of these pressure gradient parameters reflect the local direct impact of the pressure force on the boundary layer, which is the only effect experienced by \color{black} idealized equilibrium TBLs and the dominant one in near-equilibrium TBLs. \color{black}


		\begin{figure*}[h!]
			\centering
			\vspace{-0.5cm}

	\begin{tikzpicture}
	\node(a){ \includegraphics[scale=0.6]{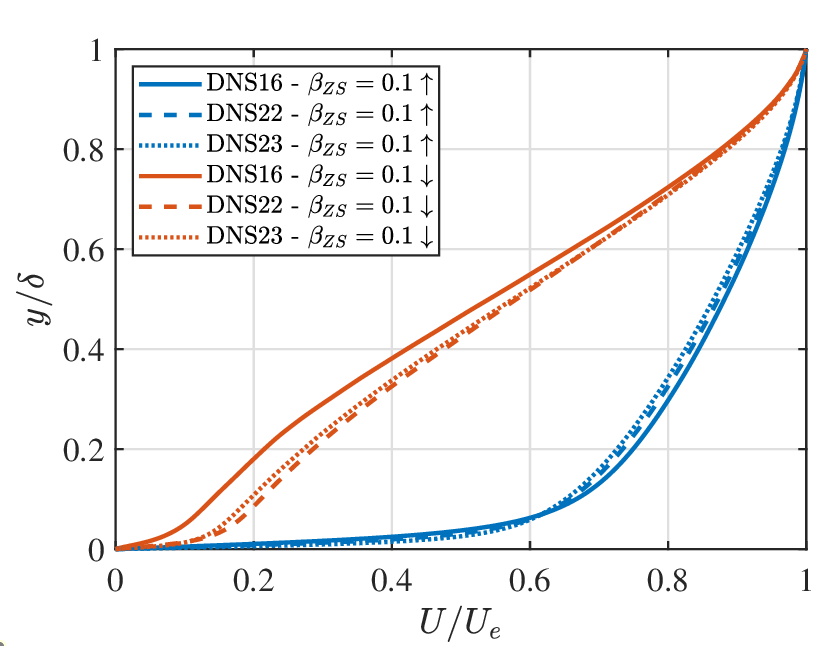}};
	\node at (-3.9,2.8) {\large $(a)$};
	\end{tikzpicture} 	\begin{tikzpicture}
	\node(a){ \includegraphics[scale=0.6]{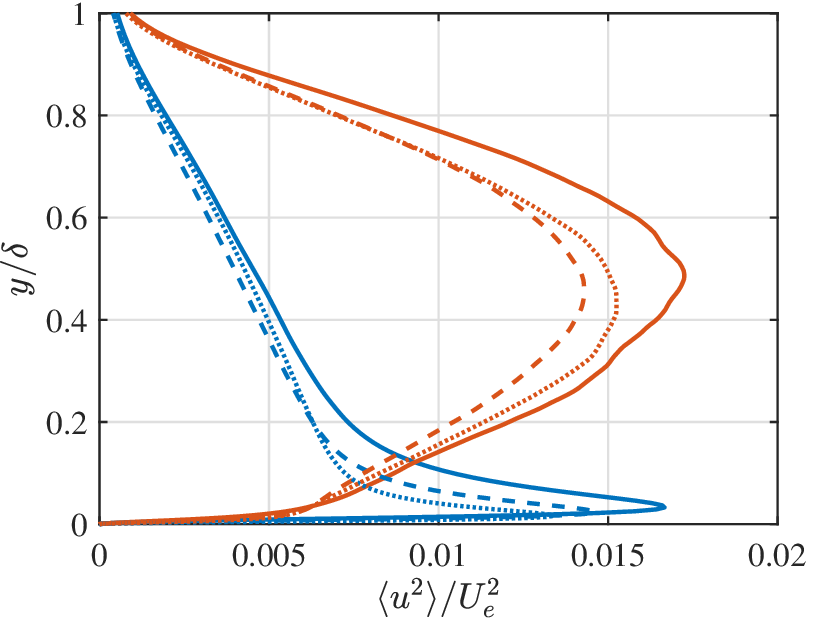}};
		\node at (-3.9,2.8) {\large $(b)$};
\end{tikzpicture}
			
			\caption{The outer-scaled mean velocity (a) and $\langle u^2\rangle$ (b) profiles of the six streamwise positions with the same $\beta_{ZS} = 0.1$ value indicated in Table 1. The arrow in the legend indicates the change rate of $\beta_{ZS}$. }
			
			\label{same_betazs}
		\end{figure*}

\color{black} The second pressure-gradient effect influencing the flow is the local disequilibrating effect, defined as the local streamwise rate of change in the relative importance of the pressure force within the overall force balance.
\color{black} We investigate this effect using the \color{black} normalized \color{black} derivative of the corresponding PG parameters in the streamwise direction: $d\beta/dX$, where $X=x/L$ and $L$ is the reference streamwise length. Unfortunately, as discussed by \cite{maciel2018outer}, $L$ cannot be easily \color{black} defined for the outer layer \color{black} if one wants to compare non-equilibrium TBLs evolving on flat plates. We choose $L$ as the average boundary layer thickness over the whole domain ($\delta_{av}$) in the outer layer, as in \cite{maciel2018outer} and as the friction-viscous length scale ($l^+ = \nu/u_\tau$) in the inner layer. 	
		
		\color{black}
		It is important to recognize that the local and cumulative (history) disequilibration effects are inherently difficult to decouple. In practice, when $d\beta/dX$ is large, the upstream cumulative effect of the pressure force is also often—but not necessarily—non-negligible. However, the design of specific flow cases to attempt this decoupling is conceivable. For instance, one could consider a near-equilibrium turbulent boundary layer ($\beta =const$, high Re) in which a finite value of $d\beta/dX$ is suddenly imposed. Observing the flow response at the location of this change would allow for the isolation of the local disequilibrating effect.
		
To clearly distinguish the local from the cumulative disequilibration effect in non-equilibrium TBLs such as those considered in the present study, one would need to define a parameter that characterizes the cumulative effect. However, this poses a formidable challenge due to the complex, nonlinear attenuation of the PG effect with distance and the delayed response of the mean flow and turbulence. While Vinuesa et al. (2017) and Gomez and McKeon (2025) have proposed accumulated PG parameters in the form of integrals to represent the cumulative effects, these formulations remain insufficiently validated and too simplistic to capture the complexities outlined above. For this reason, we have chosen not to employ a cumulative disequilibration parameter in the present study.

Figure 1 shows the spatial evolution of the main flow parameters as a function of $x/\delta_{av}$ for the three non-equilibrium TBLs, while figure 2 presents the streamwise evolution of the wall-normal profiles of $U$ and $\langle u^2 \rangle$. The profiles in figure 2 correspond to equidistant streamwise positions and are restricted to the region analyzed in this study ---that is, between the first and last symbols shown for each flow in figure 1. Starting with figure 1a, it can be seen that the three flow cases exhibit a similar spatial evolution of $\beta_{ZS}$. 	\color{black} The parameter $\beta_{ZS}$, which reflects the local direct impact of the pressure force on the outer layer, initially rises and subsequently falls. An increase of $\beta_{ZS}$ represents a momentum-losing effect of the PG in the outer region, although such an effect is not instantaneous due to the delayed response of the mean flow \citep{gungor2024turbulent}. Conversely, a decrease of $\beta_{ZS}$ signifies a momentum-gaining effect. $\beta_{ZS}$ increases faster in DNS16, denoting a stronger disequilibrating effect of the PG. 

\color{black} 
These various effects are clearly illustrated by the outer-scaled $U$ profiles shown in figure 2. The important momentum loss (i.e., increased velocity defect) is evident in all three flows, with the strongest effect observed in DNS16. For DNS23, the momentum gain associated with the decrease in $\beta_{ZS}$ is visible in the last five $U$ profiles, where the recovery initiates in the inner region. Consistent with these momentum changes, the $\langle u^2 \rangle$ profiles in figure 2 show an increase in outer-layer turbulence with increasing $\beta_{ZS}$, accompanied by a reduction in inner-layer turbulence and the eventual disappearance of the inner peak in $\langle u^2 \rangle$. In DNS23, the reduction of $\beta_{ZS}$ in the last two-thirds of the domain leads to a recovery of $\langle u^2 \rangle$ in the inner layer, and a slower decay in the outer layer, attributed to the delayed turbulence response in that region. The shape factor $H$ distributions shown in figure 1c provide a global perspective on the momentum changes. 	\color{black} As anticipated, the rise in the shape factor $H$ correlates with the intensity of the change in $\beta_{ZS}$ in the streamwise direction, $d\beta_{ZS}/dX$. The delayed response of the mean flow is also evident in this figure, as $H$ increases or decreases downstream of the corresponding increase or decrease in $\beta_{ZS}$.

Figure \ref{figure1}(b) presents the $|\beta_i|$ distributions as a function of $x/\delta_{av}$. The $\beta_i$ values vary considerably across all cases, necessitating the scaling of the $y$-axis on a logarithmic scale. The effect of the pressure force on the inner layer is clearly different from the one in the outer layer given by $\beta_{ZS}$. Whereas the relative importance of the pressure force with respect to the turbulent force begins to decline in the outer layer even before the middle of the domain, it keeps increasing in the inner layer much longer. This highlights the necessity of conducting separate analyses for the inner and outer layers. The skin-friction ($C_f$) development shown in figure \ref{figure1}(d) reveals that DNS16 has a tiny separation region whereas the other two flows exhibit no mean separation, although DNS22 almost separated. In this study, we focus on the region where there is no separation and therefore we discard the separation and re-attachment region of DNS16.

		\begin{table}[t]
			\centering
			\caption{The main properties of cases with same $\beta_{ZS}$ }
			\label{Table1}
			
			\vspace{-0.4cm} 
			\begin{tabular}{r r r r r }
				& & \\ 
				\hline
				\hline
				Name	& $\beta_{ZS}$	& $d\beta_{ZS}/d(x/\delta_{av})$	&  $Re_{ZS}$ &   \\
				\hline
				DNS16 - $\beta_{ZS}$ $\uparrow$	& 0.10  & 0.197 & 220      \\
				DNS22 - $\beta_{ZS}$ $\uparrow$	& 0.10  &   0.153   &361   \\
				DNS23 - $\beta_{ZS}$ $\uparrow$& 0.10   	& 0.035 & 588   \\
				DNS16 - $\beta_{ZS}$ $\downarrow$ & 0.10  & -0.032 &  4692  \\
				DNS22 - $\beta_{ZS}$ $\downarrow$& 0.10	&  -0.020 &  5712 \\
				DNS23 - $\beta_{ZS}$ $\downarrow$& 0.10	&  -0.023 &  10290\\
				\hline
				\hline
			\end{tabular}
		\end{table}
		

		\begin{figure*}
			\centering
			
	\begin{tikzpicture}
	\node(a){ \includegraphics[scale=0.6]{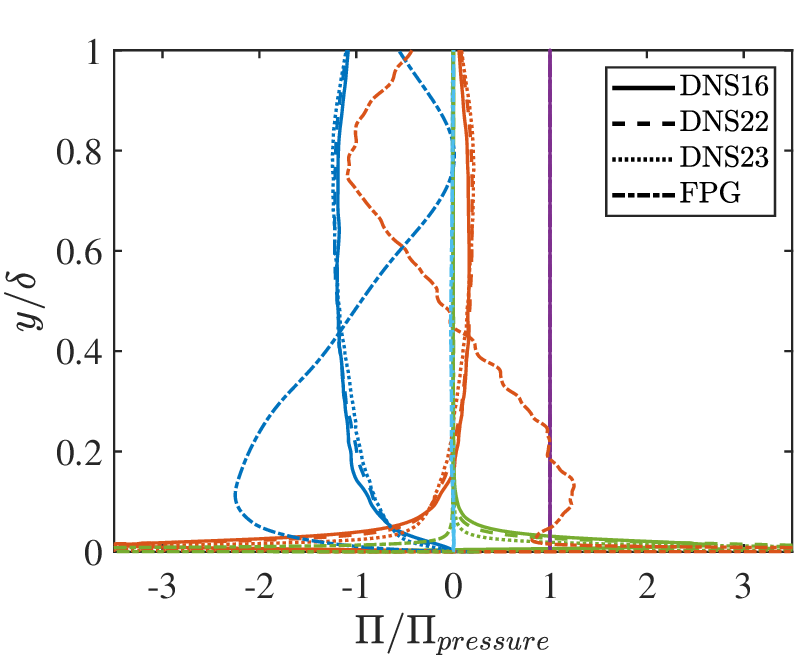}};
	\node at (-3.9,2.8) {\large $(a)$};
\end{tikzpicture} 	\begin{tikzpicture}
	\node(a){ \includegraphics[scale=0.6]{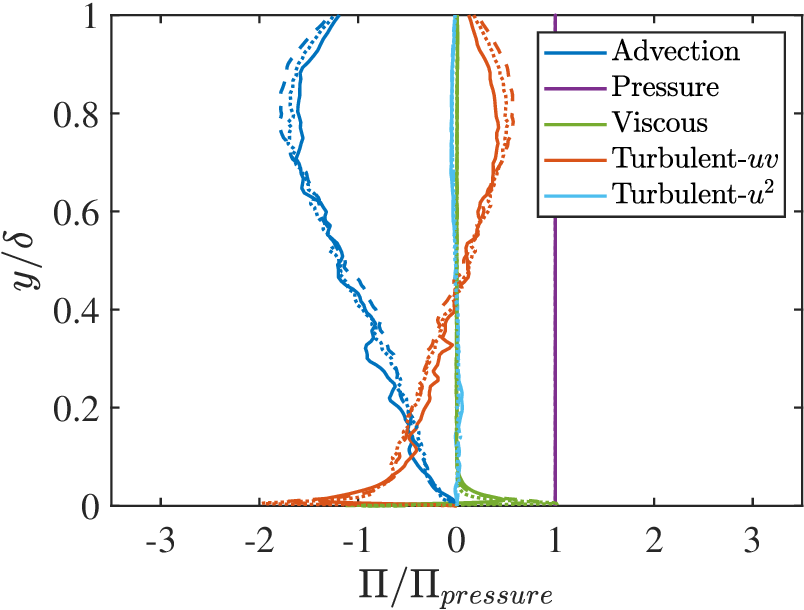}};
	\node at (-3.9,2.8) {\large $(b)$};
\end{tikzpicture}
			\caption{The terms of the mean momentum budget profiles of the small defect ($H=1.57$) (a) and large defect ($H=2.56$) cases (b) as a function of $y/\delta$ for DNS16 (straight), DNS22 (dashed) and DNS23 (dotted) and the FPG case (dashed-dotted). }
			
			\label{mom_buds}
		\end{figure*}

		\section{Results}

		\subsection{Outer layer}

		We start investigating the outer layer by comparing the three non-equilibrium flows at two streamwise positions with the same $\beta_{ZS}$ value, $\beta_{ZS}=0.1$, in the increasing-$\beta_{ZS}$ and decreasing-$\beta_{ZS}$ zones. These six streamwise positions are marked in figure \ref{figure1}a with black cross marks. This value of $\beta_{ZS}$ was selected to enable comparison between cases at the beginning of the domain, where flow history is still small. The main parameters of these cases are listed in Table \ref{Table1}. In the cases where $\beta_{ZS}$ increases (momentum-losing effect), the increase is more pronounced for DNS16 and much milder for DNS23. Conversely, in the three cases where $\beta_{ZS}$ decreases (momentum-gaining effect), the decrease is small and similar for all three flows.

		Figure \ref{same_betazs}a presents the outer-scaled mean velocity profiles of the six cases as a function of $y/\delta$. The first three flow cases (black lines in the figure) provide an opportunity to see the effect of $d\beta_{ZS}/dX$ with minimal history effects, since they are near the beginning of their respective flow domain. The difference between the three mean velocity profiles follows the expected trend: the defect increases from DNS16 to DNS23 because the mean flow cannot respond instantaneously to the increase in the pressure force. Specifically, the rapid increase in $\beta_{ZS}$ for DNS16 means that the mean flow has had less time to react compared to the other two flows. It's worth noting that the Reynolds number effect (lower $Re_{ZS}=(\delta U_{ZS}/\nu)(U_{ZS}/U_e)$ for DNS16 and DNS22 compared to DNS23) tends to reduce these differences, as the mean velocity profiles become fuller at higher Reynolds numbers. As shown in figure \ref{same_betazs}b, the $\langle u^2\rangle$ profiles are similar in the outer region and the trend is not conclusive as the Reynolds number plays an important role for the Reynolds stresses. 
		
		In the three downstream cases (orange lines in figure 3), $\beta_{ZS}$ remains identical, and $d\beta_{ZS}/dX$ is similar. This allows for the almost complete isolation of the flow history effect, which is significant in these cases, but Reynolds number effects are also present. Figure 3a illustrates that the cumulative effect of the APG results in a significant velocity defect in these three cases. DNS16 exhibits the largest velocity defect, partly due to higher upstream $\beta_{ZS}$ values and partly because of the lower Reynolds number. The differences between DNS22 and DNS23 are small as they have more similar upstream histories and Reynolds numbers. Nonetheless, the larger velocity defect for DNS23 suggests a stronger upstream impact of the PG.

		\begin{figure*}[t!]
			\centering

	\begin{tikzpicture}
	\node(a){ \includegraphics[scale=0.6]{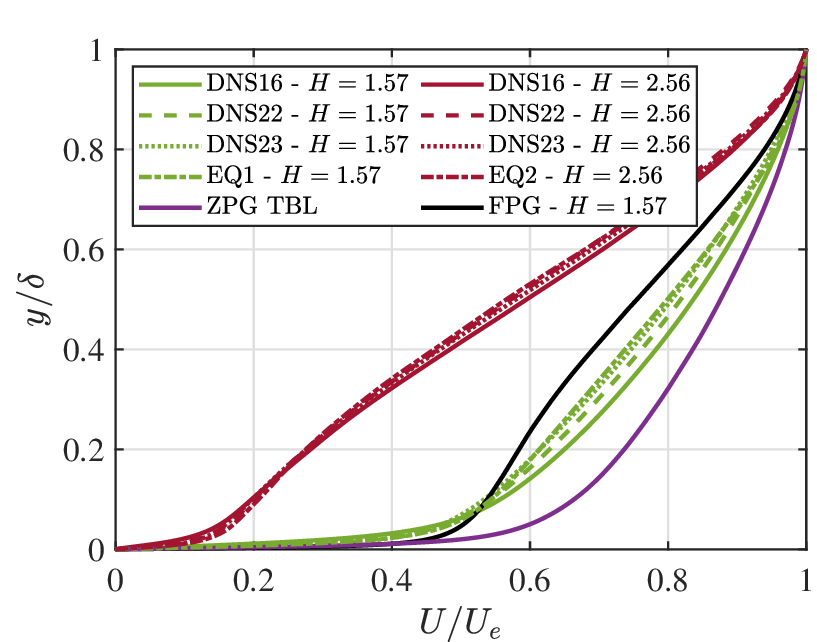}};
	\node at (-3.9,2.8) {\large $(a)$};
\end{tikzpicture} 	\begin{tikzpicture}
	\node(a){ \includegraphics[scale=0.6]{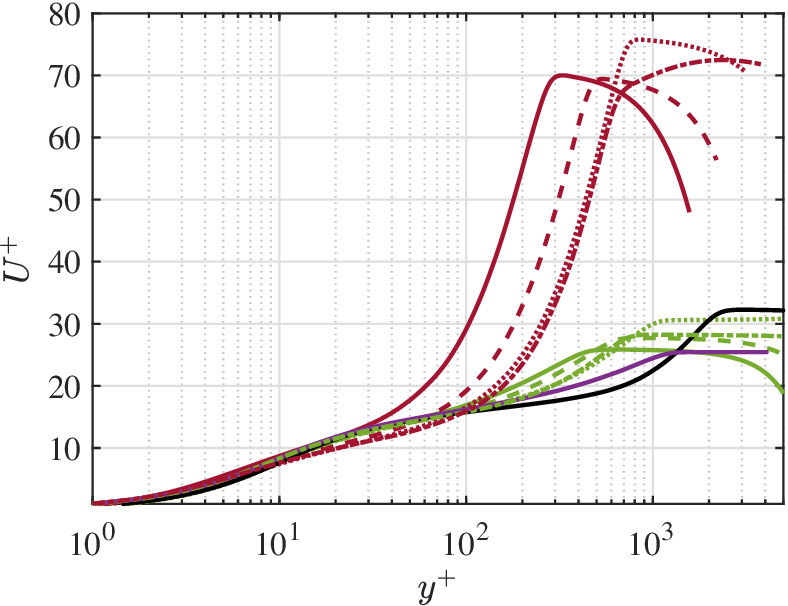}};
	\node at (-3.9,2.8) {\large $(b)$};
\end{tikzpicture}

	\begin{tikzpicture}
	\node(a){ \includegraphics[scale=0.6]{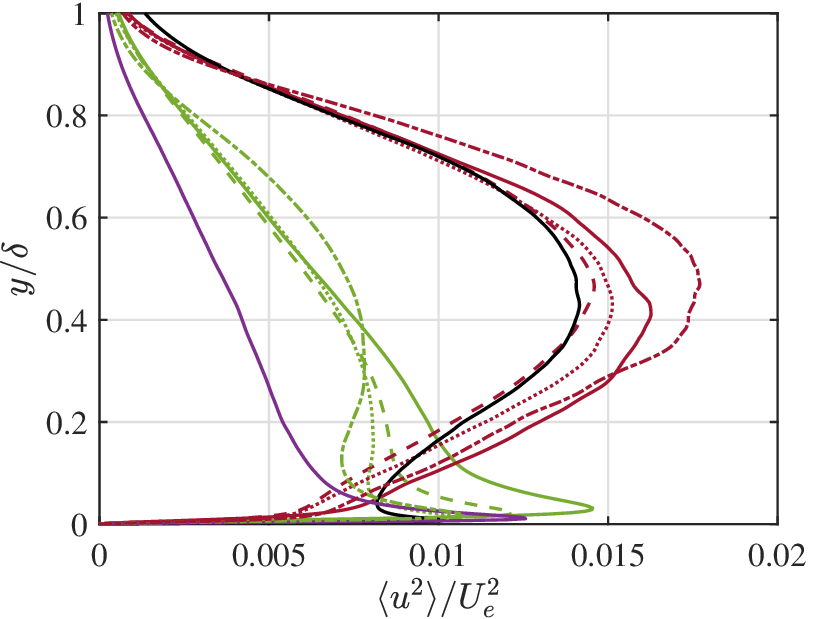}};
	\node at (-3.9,2.8) {\large $(c)$};
\end{tikzpicture} 	\begin{tikzpicture}
	\node(a){ \includegraphics[scale=0.6]{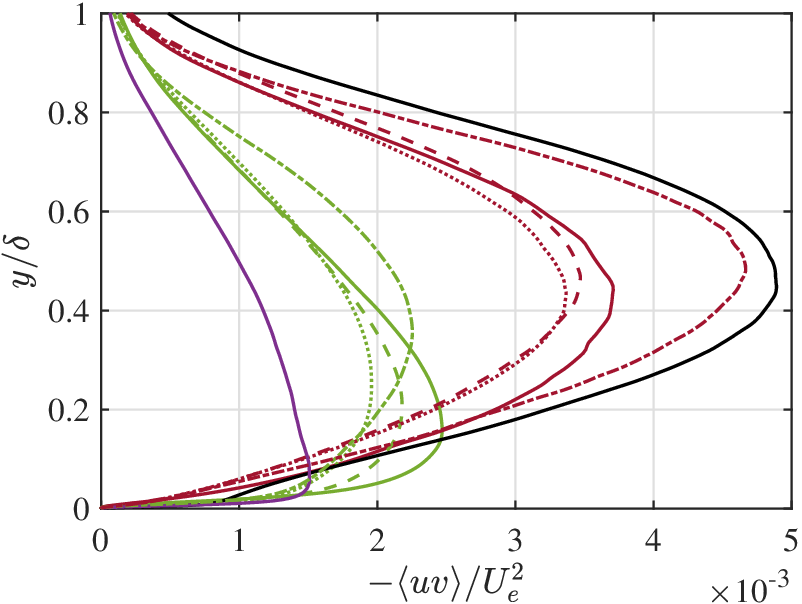}};
	\node at (-3.9,2.8) {\large $(d)$};
\end{tikzpicture}
			
			\caption{The outer- and inner-scaled mean velocity (a,b), $\langle u^2\rangle$ (c), and $\langle uv\rangle$ (d) profiles of the streamwise positions with the same shape factor, presented in Table 2, along with the ZPG TBL case of \cite{sillero2013one} as a function of $y/\delta$. }
			
			\label{same_h}
		\end{figure*}
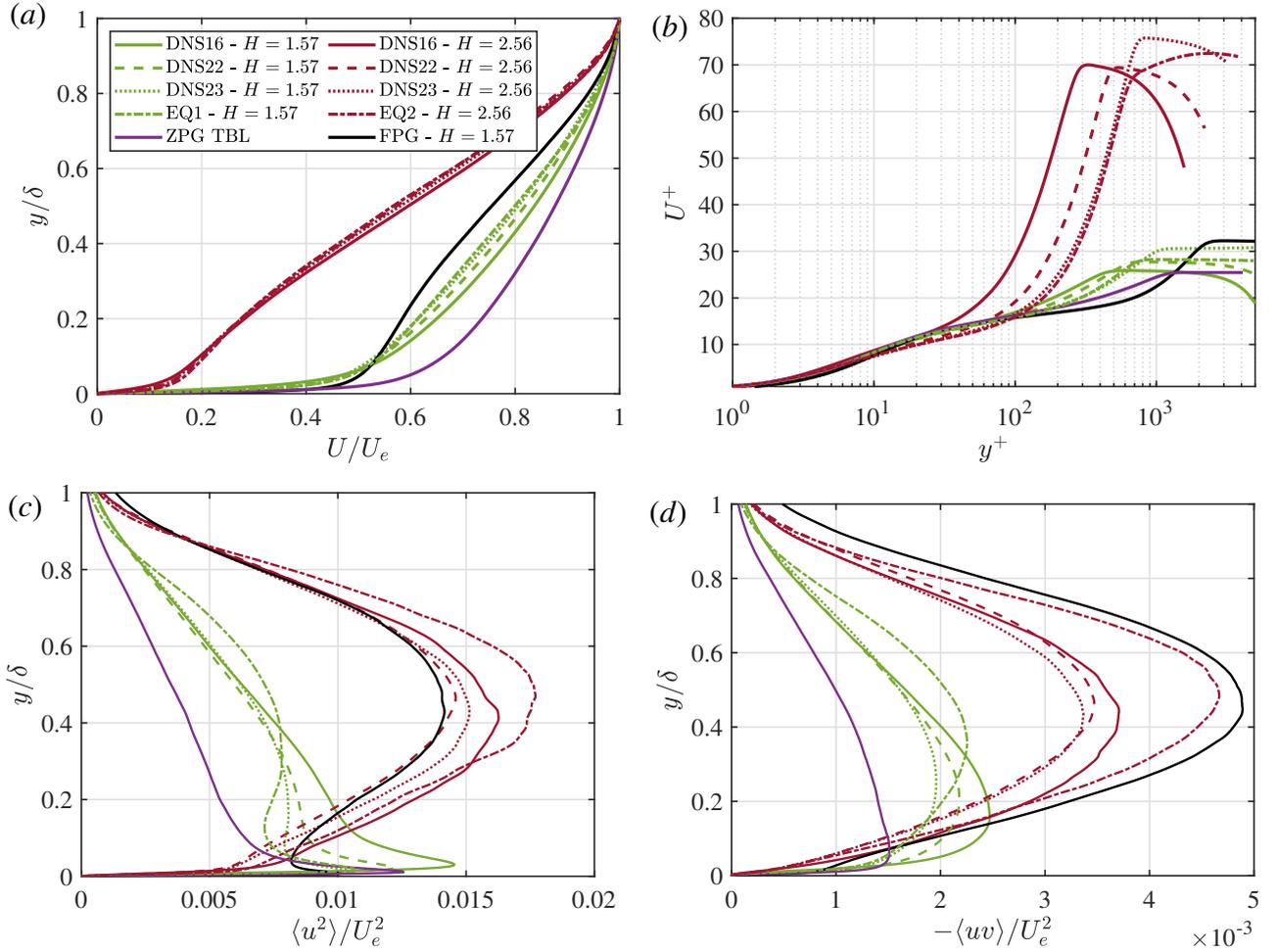

		As for the $\langle u^2\rangle$ profiles for these three cases, they are typical of strong APG cases with important outer layer turbulence and the absence of an inner peak. The higher levels of $\langle u^2\rangle$ for DNS16 are expected due to the strong upstream impact of the APG, while the lower Reynolds number may also contribute. The observation that $\langle u^2\rangle$ levels are higher for DNS23 than DNS22 is consistent with the slightly larger mean velocity defect of DNS23, again  suggesting a stronger upstream impact of the APG.

		For a different perspective, we also compare the APG TBLs at the same shape factor as it takes the cumulative effect of the pressure gradient into account. We choose two values of the shape factor, $H=1.57$ and 2.56, which correspond to the values of the near-equilibrium TBLs of \cite{kitsios2017direct}, to incorporate these cases into our analysis. By doing so, we can compare the non-equilibrium TBLs with APG TBL cases where the influence of flow history is minimal. The mean velocity defect is small for $H=1.57$ and large for $H=2.56$. We also have another position further downstream in DNS23 (FPG) where $H$ also equals 1.57, but with a favorable pressure gradient (both $\beta_{ZS}$ and $d\beta_{ZS}/dX$ are negative). The selected positions for the non-equilibrium cases are marked with orange circles for the APG cases and a black square for the FPG case in figure 1(a,c). The main properties of these cases are given in Table \ref{Table2}.

		\begin{table}[h!]
			\centering
			\caption{The main properties of cases with same $H$. }
			\label{Table2}
			\vspace*{-0.4cm}
			
			\begin{tabular}{r r r r r }
				& & \\ 
				\hline
				\hline
				Name & H	& $\beta_{ZS}$	& $d\beta_{ZS}/d(dx/\delta_{av}) $	& $Re_{ZS}$    \\
				\hline
				DNS16 	& 1.57 & 0.53     & -0.005 & 488      \\
				DNS22 	& 1.57 & 0.37  	  & -0.020 &  839    \\
				DNS23  	& 1.57 & 0.33     & 0.006 &  1622    \\
				EQ1  	& 1.57 & 0.12     & $\approx$ 0 &  1151    \\
				FPG  	& 1.57 & -0.20     & -0.049 &  4616    \\
				DNS16 	& 2.56 & 0.14     & -0.049 &  3433   \\
				DNS22  	& 2.56 & 0.09     & -0.019 &  6029   \\
				DNS23  	& 2.56 & 0.10     & -0.024 &  10051  \\
				EQ2  	& 2.56 & 0.08     & $\approx$ 0 & 8488   \\
				\hline
				\hline
			\end{tabular}
		\end{table}

		We begin by analyzing the force balance for the non-equilibrium TBLs. The near-equilibrium TBLs cannot be included due to the unavailability of their momentum budget data. Figure \ref{mom_buds} presents the outer-scaled mean momentum budgets. The momentum budget terms are normalized with the pressure force to facilitate comparison between the two defect situations. Despite different upstream histories among the three databases, the wall-normal distribution of the force balance at a given $H$ value is strikingly similar, showing near-collapse. This suggests that the force balance can be directly correlated with $H$ when both $\beta_{ZS}$ and $d\beta_{ZS}/dX$ are similar, and when the upstream flow history is similar in trend, as it is the case for the two triads of APG cases considered here. However, the mean momentum budget for the FPG case, illustrated in figure 4a, clearly shows that this is no longer true when the local pressure force and upstream history are completely different.

		\begin{figure*}[h!]
			\centering
	\begin{tikzpicture}
	\node(a){ \includegraphics[scale=0.6]{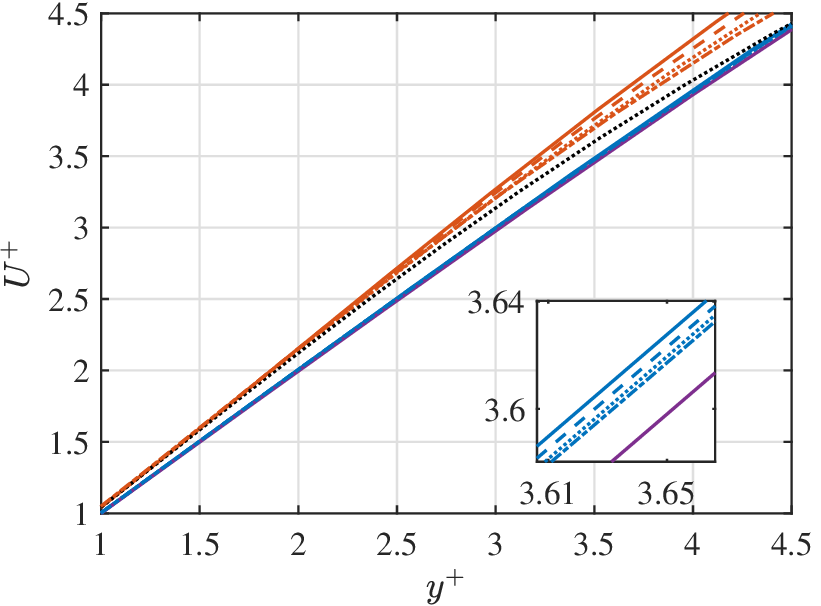}};
	\node at (-4.2,2.8) {\large $(a)$};
\end{tikzpicture} 	\begin{tikzpicture}
	\node(a){ \includegraphics[scale=0.6]{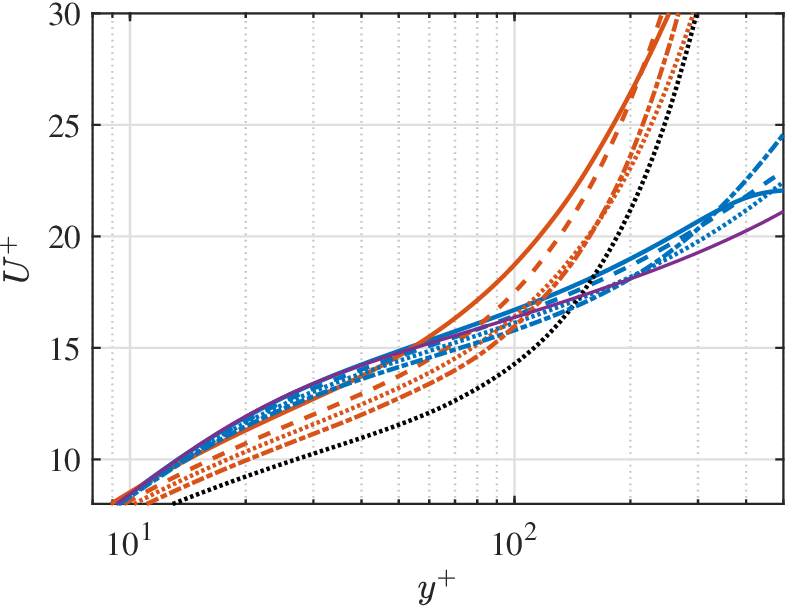}};
	\node at (-4.2,2.8) {\large $(b)$};
\end{tikzpicture}

	\begin{tikzpicture}
	\node(a){ \includegraphics[scale=0.6]{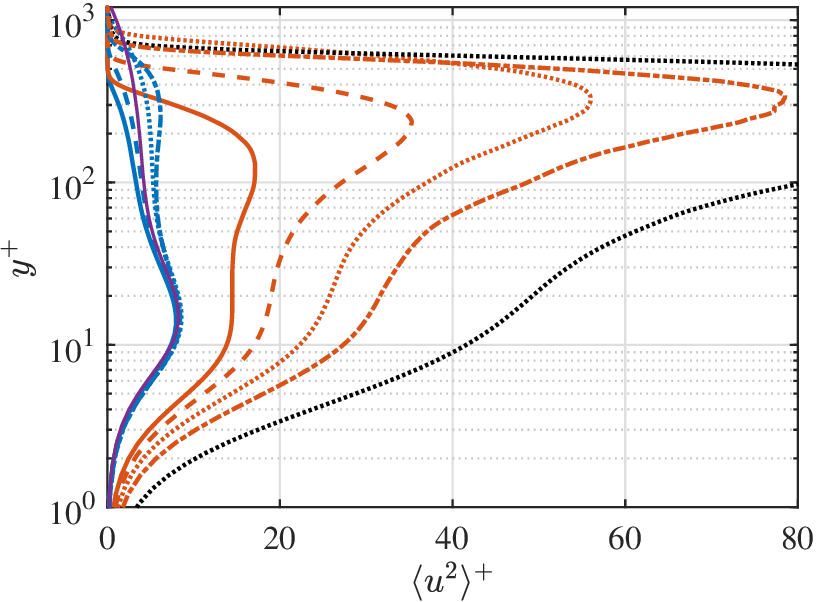}};
	\node at (-4.2,2.8) {\large $(c)$};
\end{tikzpicture} 	\begin{tikzpicture}
	\node(a){ \includegraphics[scale=0.6]{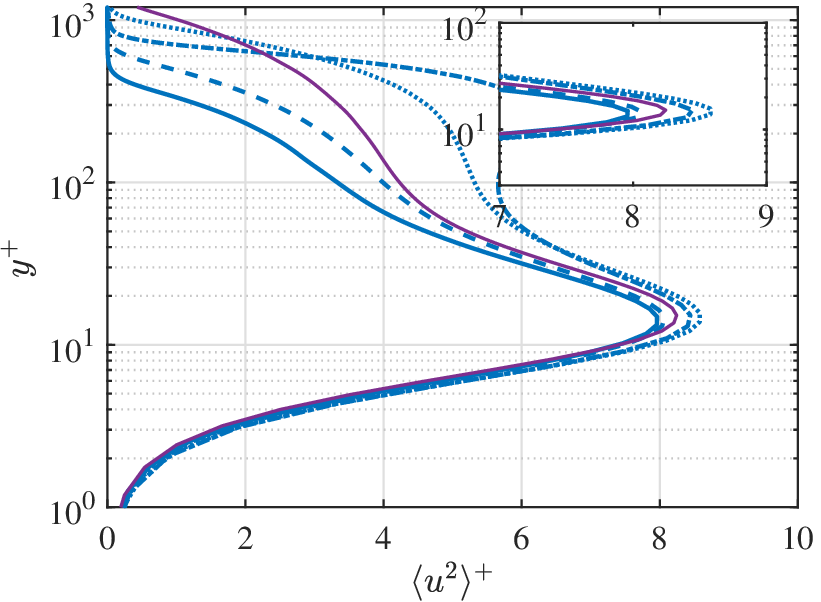}};
	\node at (-4.2,2.8) {\large $(d)$};
\end{tikzpicture}

	\begin{tikzpicture}
	\node(a){ \includegraphics[scale=0.6]{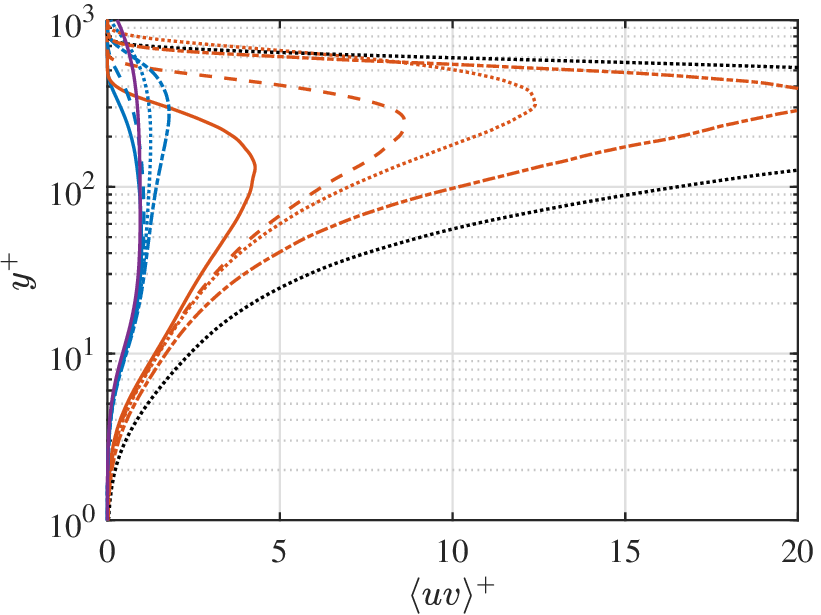}};
	\node at (-4.2,2.8) {\large $(e)$};
\end{tikzpicture} 	\begin{tikzpicture}
	\node(a){ \includegraphics[scale=0.6]{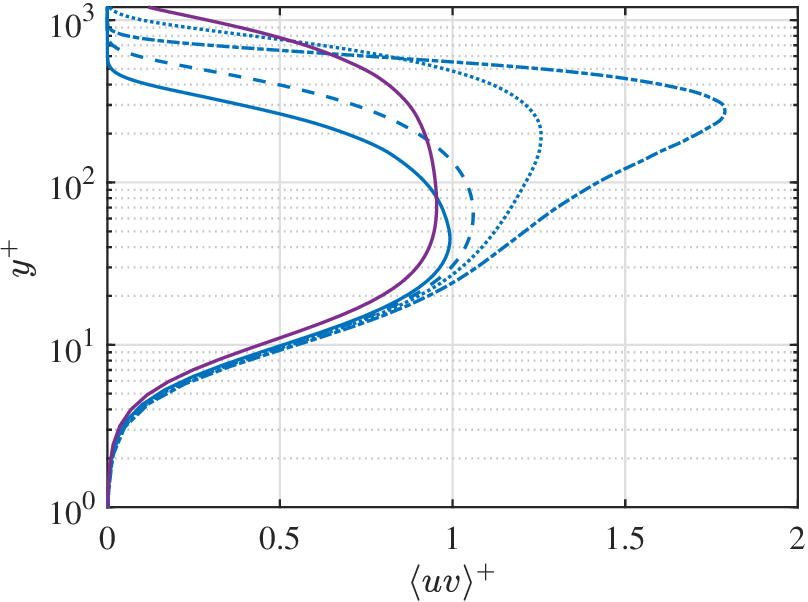}};
	\node at (-4.2,2.8) {\large $(f)$};
\end{tikzpicture}

			\caption{The friction-viscous scaled mean velocity in the viscous sublayer and overlap layer (a,b), $\langle u^2\rangle$ (c,d) and $\langle uv\rangle$ (e,f) profiles of the streamwise positions with the same $\beta_i$ value presented in Table 3. The legend is given in Table 3.}
			
			\label{upyp}
		\end{figure*}
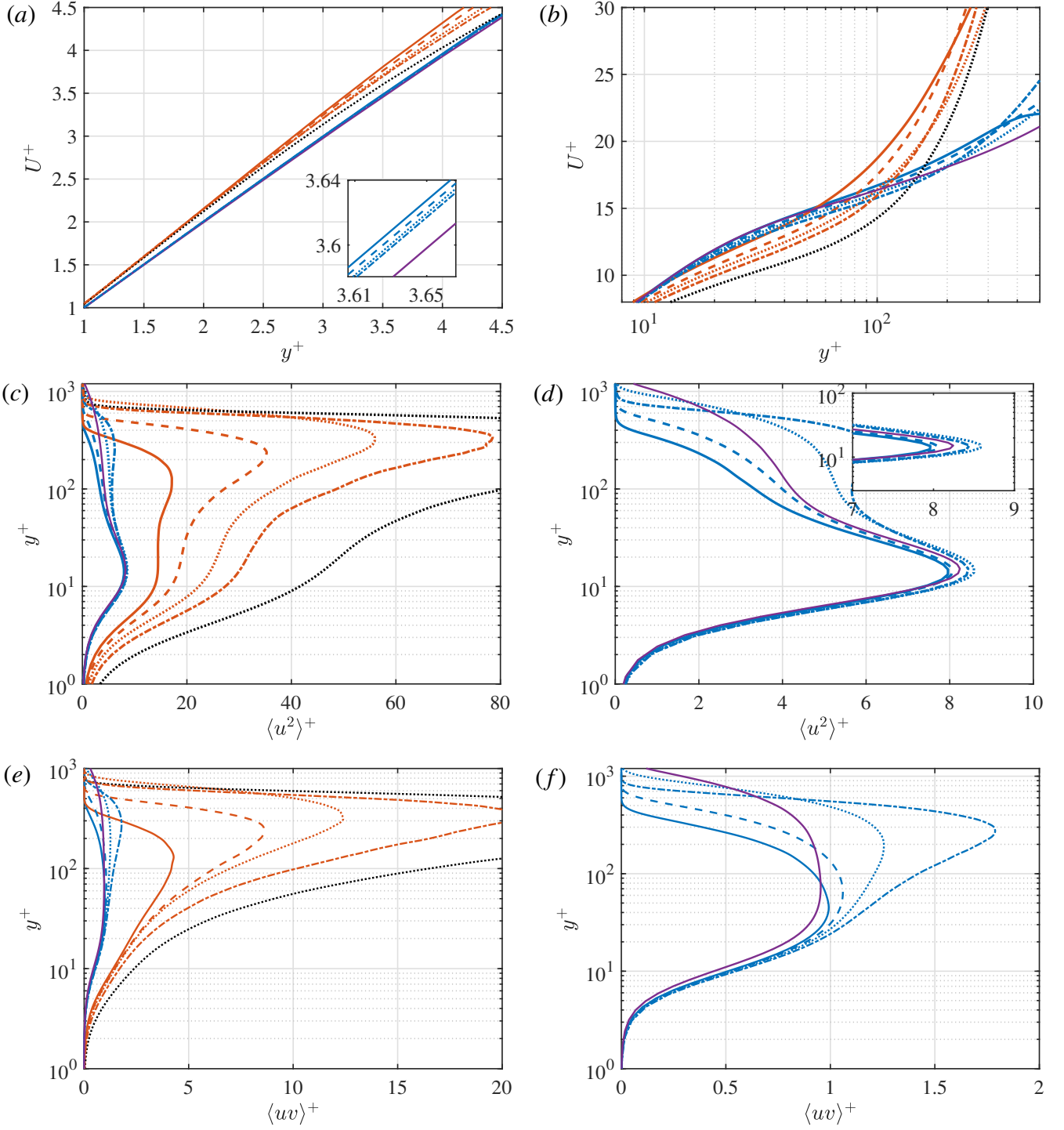

The mean velocity profiles are depicted in figure \ref{same_h}(a,b), alongside the two near-equilibrium TBLs (EQ1 and EQ2), and the ZPG TBL ($Re_\tau=1306$ and $Re_{ZS}= 14520$) from \cite{sillero2013one}. We also include the inner-scaled mean velocity profiles for reference.

Beginning with the four APG TBLs exhibiting a small defect ($H=1.57$; indicated by the green lines), we observe that while the profiles are grouped together, they are not identical, despite being APG cases with the same shape factor. The trend indicates an increasing momentum defect in the outer layer as we progress from the stronger disequilibrium case DNS16 to the near-equilibrium case EQ1. This trend is primarily attributed to the disequilibrium APG history of the flows, although it may also be partly influenced by the varying local impact of the pressure gradient, as indicated by the $\beta_{ZS}$ values in Table 2. These values decrease from DNS16 to DNS22, to DNS23, to EQ1. Additionally, the lower Reynolds number of DNS16 accentuates these differences. Since $d\beta_{ZS}/dX$ is small and similar for all non-equilibrium cases, the local destabilizing effect of the pressure gradient is likely minimal here.
		
A clear and significant effect of flow history is evident in the FPG profile at the same $H$ value (indicated by the black line). It is important to note that an equilibrium FPG TBL can never exhibit such a significant defect (a high $H$ value) as the non-equilibrium FPG case depicted in the figure. Equilibrium FPG TBLs always have fuller profiles than the ZPG TBL (purple line), regardless of $\beta_{ZS}$ value. The current FPG TBL is recovering from the pronounced upstream cumulative APG effects. As it regains momentum, it does so faster near the wall than in the outer region, resulting in a velocity profile shape markedly different from those of the four small-defect APG cases. 
		
The scenario for the four large-defect APG TBLs (indicated by the red lines) mirrors that of the small-defect cases, albeit with smaller differences and a less clear trend. In this case, the $\beta_{ZS}$ values are similar for all four TBLs.

Figure \ref{same_h}(c,d) present the $\langle u^2\rangle$ and $\langle uv \rangle$ profiles as a function of $y/\delta$ for the same APG cases discussed above. In the small-defect cases, the inner-layer turbulence is still dominant. The inner peaks of $\langle u^2\rangle$ are at higher levels than the $\langle u^2\rangle$ values in the outer layer. However, the outer-layer turbulence is elevated in all cases with respect to that in the ZPG TBL. Despite this similarity, the differences in the $\langle u^2\rangle$ profiles are more pronounced than those of the mean velocity profiles. Firstly, the broader inner peak for DNS16 reveals a significant Reynolds number effect in the lower part of this boundary layer. However, overall, the non-equilibrium TBLs do not exhibit the outer-layer plateau of the near-equilibrium TBL, between $y/\delta=0.3$ and $0.4$, even if their $\beta_{ZS}$ values are higher than that of EQ1. This indicates a delay in turbulence response to the cumulative effect of the pressure gradient, consistent with the findings of \cite{gungor2024turbulent}. Regarding the Reynolds shear stress, $\langle uv\rangle$, the profile for the near-equilibrium case EQ1 shows a maximum around $y/\delta=0.4$, whereas the non-equilibrium cases exhibit a maximum at lower $y$ values, again indicating a delay in the turbulence response.

The $\langle u^2\rangle$ and $\langle uv\rangle$ profiles of the FPG TBL show an even greater history effect in the outer layer. These profiles are similar to those of the large-defect TBLs, indicating that turbulence has not yet recovered from the upstream APG effects. In an equilibrium, FPG TBL, the Reynolds stress levels in the outer region would be lower than that of the ZPG TBL.

In the large-defect case, the inner peak of the $\langle u^2 \rangle$ profiles vanishes and the outer-layer turbulent activity becomes dominant. Differently from what we observe for the small-defect cases, the wall-normal distribution of the $\langle u^2\rangle$ profiles is very similar across the different cases. The $\langle u^2 \rangle$ levels of the non-equilibrium TBLs are nonetheless lower than that of the near-equilibrium case, indicating again a delayed response of turbulence. Similar observations can be made for the $\langle uv \rangle$ profiles. One interesting observation is that the FPG case has a higher Reynolds shear stress level than the large-defect case, which contrasts with the behavior observed for the $\langle u^2\rangle$ profiles.

		


		\subsection{Inner layer}
		
To study the inner layer response to the pressure force, we examine flows at identical values of the inner-layer PG parameter, $\beta_i$. Thus, we select positions in the non-equilibrium TBLs with the same  $\beta_i$ values as the near-equilibrium TBLs: $\beta_i=0.0066$ and $0.1000$. The streamwise positions of the non-equilibrium cases are indicated in Figure 1b with diamond symbols. They are located in a region where  $\beta_i$ increases, except for the third position in DNS23. The main parameters of these cases are listed in Table 3.

		\begin{table}[ht]
			\centering
			\caption{The main properties of cases with same $\beta_i$ }
			
			\vspace*{-0.4cm}
			
			\label{Table3}
			\begin{tabular}{r r r r r }
				& & \\ 
				\hline
				\hline
				Name	& $\beta_i$	& $d\beta_i/dx^+$ $\times$ $10^4$	& $Re_\tau$   & \\
				\hline
				DNS16 - Low $\beta_i$	& 0.0066    	& 0.0521    & 427 & \color{colmm1}{\textbf{\sampleline{}} }  \\
				DNS22 - Low $\beta_i$	& 0.0066    	&  0.0208   & 581 & \color{colmm1}{\sampleline{dashed} } \\
				DNS23 - Low $\beta_i$	& 0.0066 &  0.0068 & 957 & \color{colmm1}{\sampleline{dotted} } \\   
				EQ1 - Low $\beta_i$	   & 0.0066	& $\approx 0$ & 793 & \color{colmm1}{\sampleline{dash pattern=on .7em off .2em on .2em off .2em}}  \\
				DNS16 - High $\beta_i$& 0.1000	    & 0.9046 & 411 &  \color{colmm2}{\textbf{\sampleline{}} }\\
				DNS22 - High $\beta_i$& 0.1000	        &  0.4396 & 538 & \color{colmm2}{\textbf{\sampleline{dashed}} }   \\
				DNS23 - High $\beta_i$& 0.1000	    &  0.3365 & 800  & \color{colmm2}{\textbf{\sampleline{dotted}} }  \\
				DNS23 - High $\beta_i$& -0.1000	    & -0.6552 & 831 &  \color{black}{\textbf{\sampleline{dotted}} }   \\	
				EQ2 - High $\beta_i$	   & 0.1000	& $\approx 0$ & 691  &  \color{colmm2}{\textbf{\sampleline{dash pattern=on .7em off .2em on .2em off .2em}} }  \\
				ZPG   & 0	& $ 0$ &   1306	 &  \color{colmm4}{\textbf{\sampleline{}} }  \\
				\hline
				\hline
			\end{tabular}
		\end{table}

Figure 6a shows $U^+$ as a function of $y^+$ in the viscous sublayer for all these cases along with the ZPG TBL. Since inertia effects are minimal in the near-wall region, as expected, the profiles tend to cluster together for a given value of $\beta_i$. At the low $\beta_i$ value (0.0066), they nearly collapse, slightly above the ZPG TBL profile. However, the inset in Figure 5a, zooming in on the region around $y^+=3.6$, reveals a slight trend. Given the rapid response of the near-wall region to changes, this is likely more indicative of a local disequilibrating effect ($d\beta_i/dx^+$) than a cumulative effect. The fact that $U^+$ increases with increasing $d\beta_i/dx^+$ tends to support this hypothesis. It is essential to stress that the changes observed in the low-$\beta_i$ cases (in the inset of figure 6a) are very small and must be interpreted with care.
		
A similar trend is observed for the high $\beta_i$ cases ($\beta_i=0.1$), but it is much more pronounced. Once again, $U^+$ increases with increasing $d\beta_i/dx^+$. Additionally, the black dotted curve corresponds to negative $d\beta_i/dx^+$ and lies below the near-equilibrium curve. These observations further suggest that the variations are primarily attributable to the local disequilibrating effect of the pressure gradient, although a minor historical effect may also be present. 
		
In the overlap layer, we again employ the ZPG TBL profile and the deviation from it to assess the effect of the PG. The behaviour of the four TBL cases presented in the figure 6b are very different from what we observe in the viscous sublayer.  DNS16 profiles have the smallest deviation from the ZPG TBL profile for both small- and low-$\beta_i$ cases and the case with negative $d\beta_i/dx^+$ has the highest deviation. The behavior in the overlap region is challenging to interpret, as it is influenced by the responses of the inner and outer layers, which differ significantly.

		\begin{figure*}[h!]
			
			\centering
			
			\includegraphics[scale=0.5]{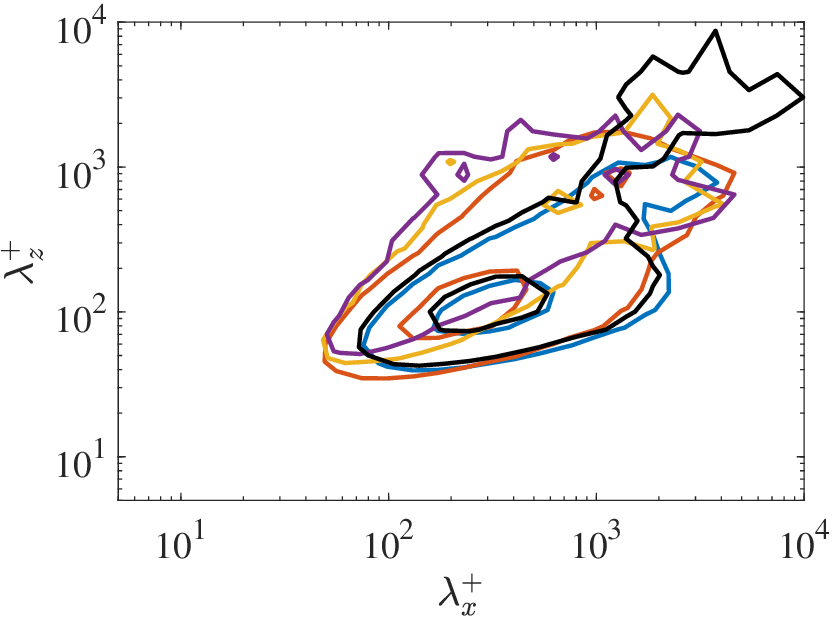}
			\includegraphics[scale=0.5]{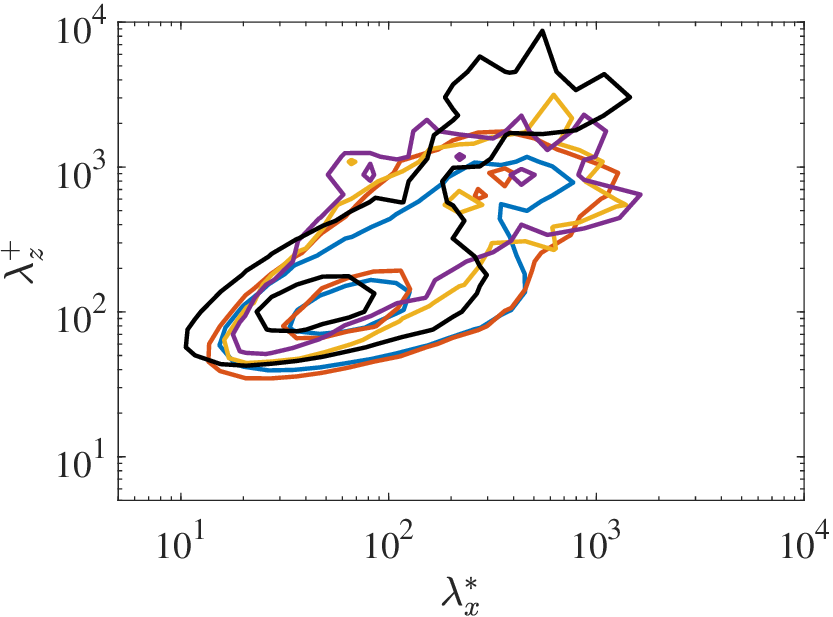}
			\includegraphics[scale=0.5]{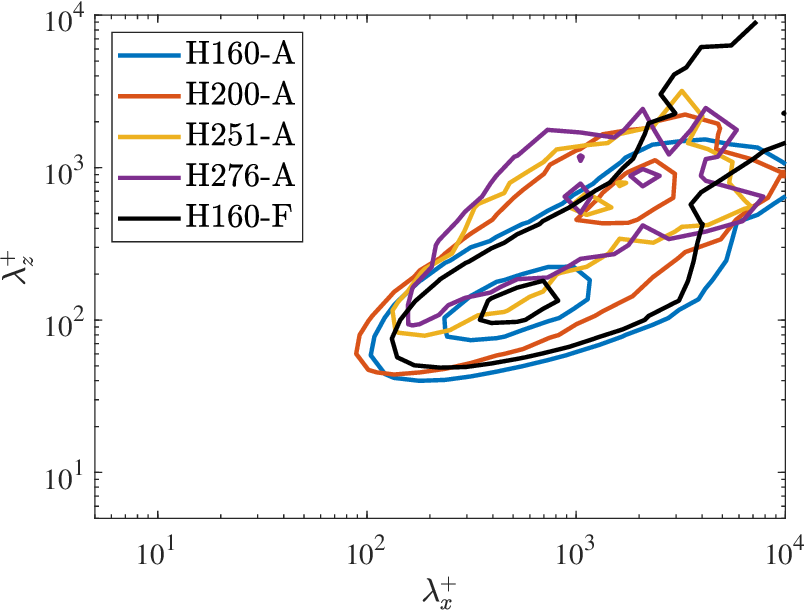}
			\includegraphics[scale=0.5]{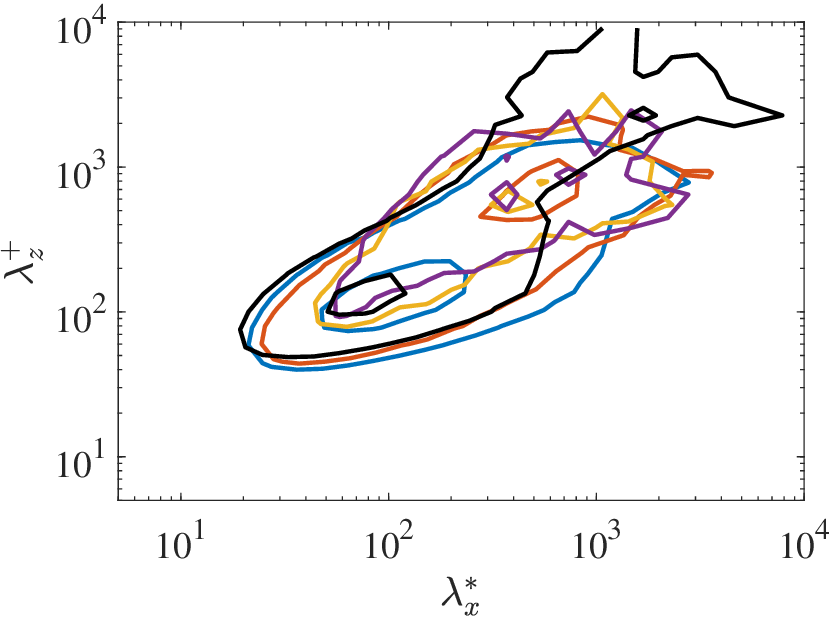}

			\caption{ The two-dimensional energy spectra of $\langle u^2\rangle$ as a function of $\lambda_x^+$ and $\lambda_z^+$ (left) and  $\lambda_x^*$ and $\lambda_z^+$ (right) at $y^+=5$ (top) and $y^+=15$ (bottom) for DNS23. The lines indicate the streamwise positions given in the legend and Table 4. The contour levels are [0.8 0.5] of the maxima of spectra}
		\label{2dspec_uu}
			
		\end{figure*}

Figure 6(c,d) shows the $\langle u^2\rangle^+$ profiles as a function of $y^+$. Figure 6(d) displays a smaller range of $\langle u^2\rangle^+$ values to highlight the behavior of the small-defect (low-$\beta_i$) cases. The inner peak for the $\langle u^2\rangle$ profiles in the low-$\beta_i$ cases exists in all four cases and the shape of the profiles in the inner layer are almost identical even though the outer layers behave differently from each other. The $\langle u^2\rangle$ profile's peak is located at $y^+\approx 14-15$ but there is a minor difference between the levels. Regarding this difference, we first focus on the near-equilibrium TBLs (ZPG and EQ1) to see the effect of $\beta_i$. The peak of $\langle u^2\rangle^+$ is higher in EQ1 than in the ZPG TBL, despite EQ1 having a considerably lower $Re_\tau$ compared to the ZPG TBL. Given that the peak level is known to increase with Reynolds number in the case of the ZPG TBL, these results suggest that positive values of $\beta_i$ also contribute to an increase in the inner peak of $\langle u^2\rangle^+$. Now, when we compare the inner peaks of the four APG TBLs at the same low $\beta_i$ value, the variations do not align consistently with the changes of $d\beta_i/dx^+$. Instead, they correspond with variations in $Re_\tau$. However, the effect of $d\beta_i/dx^+$ cannot be discarded because we cannot isolate any of these factors with the existing cases.  Nonetheless, the difference in levels is small. Overall, the similarity of the $\langle u^2\rangle$ profiles in the inner layer suggests that flow history plays a minor role in these cases.
		
The picture changes significantly in the high-$\beta_i$ cases. The Reynolds stress levels are very different between these cases, with the same trend throughout the whole boundary layer. This phenomenon can likely be attributed to the dominance of turbulence in the outer layer in large-defect TBLs \citep{gungor2016scaling, gungor2022energy}. Consequently, the observed increase in $\langle u^2\rangle$ from the stronger disequilibrium case DNS16 to the near-equilibrium case EQ2 probably reflects historical effects in the outer region, including the delayed turbulence response. 

For the Reynolds shear stresses, which are shown in figure 6(e,f), the levels are much higher in the high-$\beta_i$ cases than in the low- $\beta_i$ cases. In the low- $\beta_i$ cases, the differences between the four cases in the inner layer are more pronounced for $\langle uv \rangle^+$ than for $\langle u^2\rangle^+$ . This is likely due to the more significant variations in $\langle uv\rangle^+$  in the outer region compared to $\langle u^2\rangle^+$.

To examine the effects of the pressure force on inner-layer turbulent structures, we analyze their size and aspect ratio using the two-dimensional spectral distributions of $\langle u^2\rangle$ and $\langle uv\rangle$ at two wall-normal positions in the inner layer. The first one is in the viscous sublayer at $y^+=5$ and the other one is in the buffer layer at $y^+=15$ where $\langle u^2\rangle$ peaks in the small-defect cases. For simplicity, we present spectra only from DNS23. Note that the spectral distributions are computed at streamwise positions different from those previously used to analyze the inner layer. This is because we employ spatio-temporal data, which is available only at specific positions. Consequently, we cannot examine positions that isolate various effects as done earlier. Table \ref{2d_inner} summarizes the streamwise positions where the spectral distributions are computed in the inner layer. These positions are also marked with green squares in figure \ref{figure1}.

		As mentioned above, we employ spatio-temporal data (spatial in the spanwise direction) for computing the 2D spectra. We invoke Taylor's frozen turbulent hypothesis to convert the frequency into streamwise wavenumber. 
		
		\begin{equation}
			k_x = \frac{2\pi f}{U_c}
		\end{equation}
		
		\noindent 		where $f$ is the sampling frequency, $U_c$ is the convection velocity, which is considered to be the local mean velocity.

		\begin{table}
			\centering
			\caption{The main properties of cases with the 2D spectra of DNS23 flow case.  In the PG column, A stands for adverse and F for favorable. }
			\label{Table_spectra}
			
			\vspace{-0.4cm} 
			\begin{tabular}{l r r r r r}
				& & \\ 
				\hline
				\hline
				Name	& $H$	& $\beta_i$	& $d\beta_i/dx^+ \times 10^4$& $Re_{\tau}$ & PG   \\
				\hline
				H160-A 	& 1.60 & 0.0165 &0.0240& 1010  & A    \\
				H200-A  & 2.01 & 0.0552  &0.1516 &917 & A\\
				H251-A  & 2.51 & 0.1267 &0.4004& 749 & A\\
				H276-A  & 2.76 & 0.1627 &0.2298 &  700& A\\
				H160-F  & 1.60 & -0.0051 &-0.0003 &  2156 & F \\
				\hline
				\hline
			\end{tabular}
			\label{2d_inner}
		\end{table}

		Figure \ref{2dspec_uu} presents the 2D pre-multiplied energy spectra of $\langle u^2\rangle$ as a function of streamwise and spanwise wavelengths ($\lambda_x$ and $\lambda_z$) at the two wall-normal positions, $y^+=5$ and $y^+=15$. The wavelenghts are normalized using the friction-viscous length scale and a mixed length scale which is a combination of friction-viscous and pressure-viscous scales. We define the mixed length scale as follows
		
		\begin{equation}
			l^* = l^+ + l^{pi}
			\label{mixed_len}
		\end{equation}
		
		\noindent where $l^+=\nu/u_\tau$ and $l^{pi}=\nu/u^{pi}$. 
		
The inner pressure velocity scale, $u^{pi}$, is defined in equation \ref{upi}. 

			\begin{equation}
	u^{pi} = \frac{\nu}{\rho}\frac{dp}{dx}
	\label{upi}
\end{equation}

\noindent By including $l^{pi}$, we account for the local effect of the pressure force in the inner layer alongside friction forces. Similar length or velocity scale combinations have been suggested \citep{skote2002direct,manhart2008near,ma2024scaling}, but we are not aware of prior use of the mixed length scale $l^*$ in equation \ref{mixed_len}. Note that $l^*$ differs from the mixed length scale of \cite{ma2024scaling}, $\nu/(u_\tau+u^{pi})$. We tested the latter but found the results to be inconclusive, unlike for $l^*$.


It is important to stress that we employ the mixed length scale only for $\lambda_x$ and keep using the friction-viscous length scale for $\lambda_z$, because the spectra show that the pressure force does not affect the spanwise extent of structures. 

The  $\langle u^2 \rangle$ spectra on the left plots of figure 7 show that the friction-viscous length, $l^+$, scales the inner peak wavelengths for the small defect cases H160A and H160-F (black and black lines) at both $y$-positions. The inner peak reflects the streaks. This scaling suggests that flow history has little impact on the inner layer of small-defect TBLs, as H160-A and H160F have substantially different flow histories. In the large-defect cases, the spectra change shape entirely, with scale broadening caused by outer-layer turbulence influencing the inner layer, as previously reported \citep{gungor2022energy, gungor2024turbulent2}. Due to this complex behavior, wavelength scaling of the $\langle u^2\rangle$ spectra is not achieved, unlike for the $\langle uv\rangle$ co-spectra, which are now discussed.


\begin{figure*}[t!]  
	\centering 
	\includegraphics[scale=0.5]{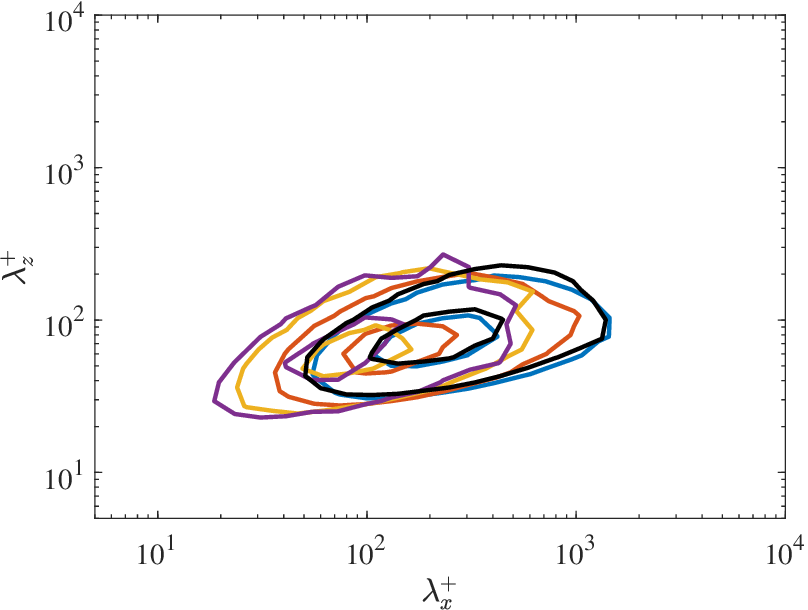}
	\includegraphics[scale=0.5]{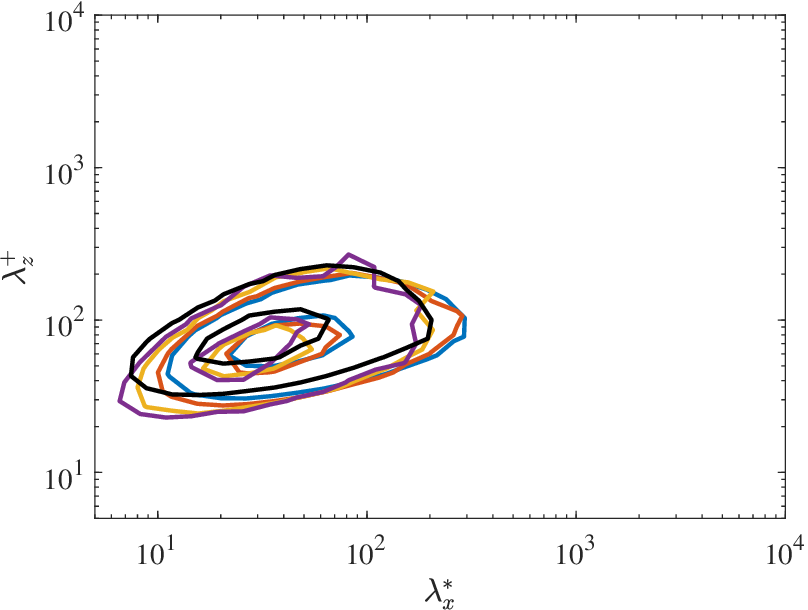}
	\includegraphics[scale=0.5]{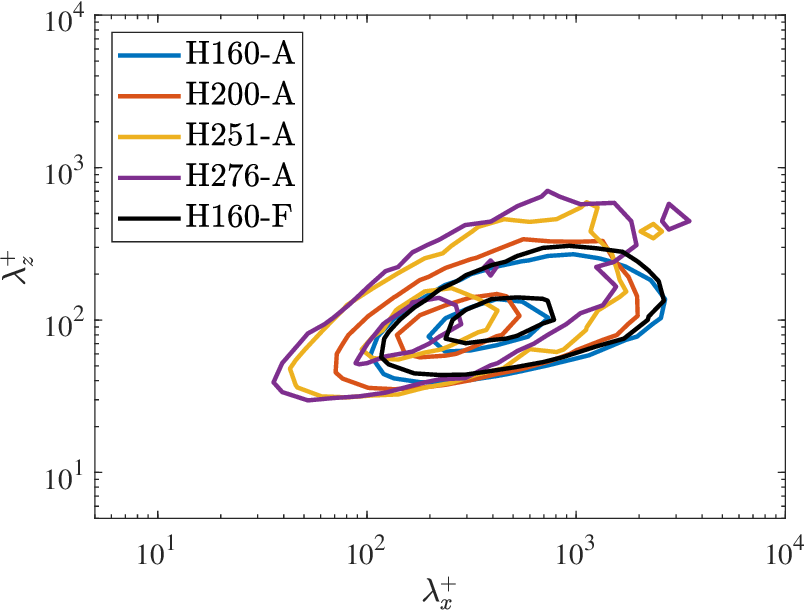}
	\includegraphics[scale=0.5]{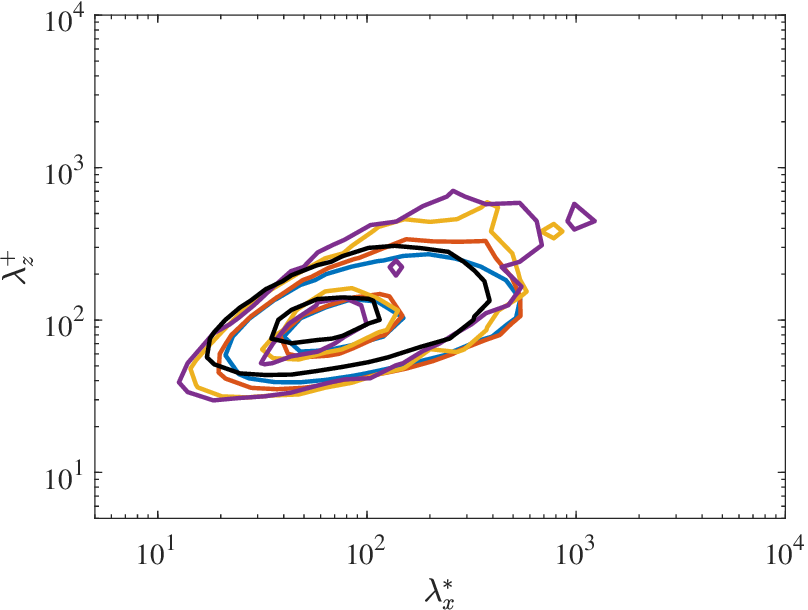}
	\caption{ The two-dimensional co-spectra of $u$ and $v$ as a function of $\lambda_x^+$ and $\lambda_z^+$ (left) and  $\lambda_x^*$ and $\lambda_z^+$ (right) at $y^+=5$ (top) and $y^+=15$ (bottom) for DNS23. The lines indicate the streamwise positions given in the legend and Table 4. The contour levels are [0.8 0.5] of the maxima of spectra.}
	\label{2dspec_uv}
\end{figure*}

 Figure \ref{2dspec_uv} shows the 2D pre-multiplied co-spectra of $u$ and $v$ at two wall-normal positions as in figure \ref{2dspec_uu}. The shape of the spectra does not change significantly even in the large-defect cases, unlike the $\langle u^2\rangle$ spectra, especially at $y^+=5$. This shows that the effect of the outer layer on the inner-layer $uv$-structures is less important than on the $u$-structures, and supports the view that the effect of upstream flow history on the inner layer is either minimal or short-lived. The friction-viscous length, $l^+$, scales the spectral wavelengths almost perfectly in the two small-defect cases but fails to collapse the streamwise wavelengths in the large-defect cases at both $y$-positions. In contrast, the mixed length scale $l^*$ successfully scales the streamwise wavelengths of the $uv$ spectra regardless of the velocity defect —an impressive result not achieved with the mixed length scale of \cite{ma2024scaling}, $\nu/(u_\tau+u^{pi})$. This finding suggests that the size of $uv$-structures in the inner layer is independent of flow history, as friction-viscous and pressure-viscous scales depend only on local variables. It also implies that $uv$-structures are largely unaffected by large-scale outer structures. Furthermore, the decrease in $\lambda_x^+$ for the $uv$ peak with increasing $\beta_i$ suggests that the pressure force compresses the $uv$-structures in the streamwise direction.

\color{black}
These differences in the $\langle u^2 \rangle$- and $\langle uv \rangle$-spectra do not imply that smaller, more localized $u$-structures—such as near-wall streaks—change substantially between small-defect and large-defect cases. \cite{gungor2024turbulent2} showed that streaks and the near-wall cycle persist in large-defect APG TBLs when outer-layer turbulence is artificially removed. However, this does not necessarily imply that streaks are present in real large-defect APG TBLs; they may be absent or occur less frequently. What is clear, however, is that the $\langle u^2 \rangle$-spectra are modified by the imprint of large-scale outer-layer $u$-structures. These energetic structures dominate the flow, making it more difficult to isolate smaller local $u$-structures within the spectra. Yet, \cite{gungor2022energy} found that the spectral imprints of outer-layer $u$-structures contribute little to local energy transfer processes. Consequently, the $uv$-structures—directly responsible for local turbulence production—are not significantly affected by these outer-layer features.

\color{black}

		\section{Concluding remarks}
		
		In this study, we explored the effect of the flow history on the inner and outer layers of turbulent boundary layers, along with the influence of the local pressure force and its variation (local disequilibrating effect).
		The main conclusions are as follows:

		\begin{enumerate}
			
			\item First, it is important to state that the five flow cases used in this study (three non-equilibrium, two near-equilibrium) do not allow for a clear-cut separation between the four effects present, namely the three pressure force effects and the Reynolds number effect. Achieving this separation would require designing and generating numerous flow cases, each with only one of these four effects changing. However, such an extensive set of flow cases does not currently exist and would be challenging to implement. Nevertheless, because they represent three distinct degrees of flow disequilibration, the three non-equilibrium TBLs analyzed in this study offer valuable insights into these three pressure force effects.
			
			\item By examining positions from the different flows with the same positive (APG) value of the pressure gradient parameter $\beta_{ZS}$, and located near the beginning of the flow domains, it becomes evident that the streamwise rate of change of the pressure gradient impact ($d\beta_{ZS}/dX$) influences the mean velocity profiles in the outer layer. Essentially, the mean flow does not respond instantaneously to the pressure force increase: the more rapid the increase, the more delayed is the response. 
			
			\item  In another comparison where $\beta_{ZS}$ was identical and $d\beta_{ZS}/dX$ was similar between flow cases, the cumulative impact of the pressure force (history effect) became apparent. A stronger cumulative APG impact resulted in a larger momentum defect and increased Reynolds stresses in the outer layer.

			\item  Even when comparing different APG TBLs with the same value of the shape factor, which necessarily implies mean velocity profiles that should be similar, these profiles differ slightly between flow cases. The results suggest that this is mainly due to the different APG cumulative effects rather than local effects. The differences are even more pronounced for the streamwise Reynolds normal stress, revealing a clear delay in turbulence response. The FPG TBL at the end of one flow, with the same shape factor as the small-defect APG cases, exemplifies a drastic flow history effect—specifically, a prolonged upstream APG. Its shape factor matching that of small-defect APG TBLs is already unusual. Its mean velocity and Reynolds stress profiles deviate significantly from those of small-defect APG TBLs and even more so from the expected behavior of FPG TBLs.
			
			\item Regarding the inner layer, a comparison of the flows at identical values of the inner pressure gradient parameter $\beta_i$ reveals that the friction-viscous scaled mean velocity profiles in the viscous sublayer are grouped according to the $\beta_i$ value. However, even in this near-wall region where inertia effects are minimal, the profiles at identical $\beta_i$ values differ slightly due to the local disequilibrating effect of the pressure gradient ($d\beta_i/dx^+$). As for the $\langle u^2\rangle$ profile in the inner layer, the effect of the pressure gradient is minor for small values of $\beta_i$ but becomes important for higher values of $\beta_i$, corresponding to a significant mean velocity defect. In the latter case, the results suggest that it is the response of outer-layer turbulence to the pressure force effects that controls what happens in the inner layer.
			
\item The latter conclusion is confirmed by the analysis of the $\langle u^2 \rangle$ spectra, which change shape entirely in the inner layer due to outer-layer turbulence. However, the situation is completely different for the $uv$ co-spectra, whose streamwise wavelengths scale with the mixed scale $\nu / u_\tau + \nu / u^{pi}$, and spanwise wavelengths scale with the friction-viscous scale $\nu / u_\tau$. Since these are local scales, the size of $uv$-structures in the inner layer is independent of flow history and large-scale outer structures.

		\end{enumerate}
		
		\section{Acknowledgments}
		
		We acknowledge the EuroHPC Joint Undertaking for awarding the project access to the EuroHPC supercomputer Leonardo DCGP at CINECA, Italy, through a EuroHPC Regular Access call, PRACE for awarding us access to Marconi100 at CINECA, Italy and Calcul Québec (www.calculquebec.ca) and
		the Digital Research Alliance of Canada (alliancecan.ca) for awarding us access to Niagara HPC server. TRG and YM acknowledge the support of the Natural Sciences and Engineering Research Council of Canada (NSERC), project number RGPIN-2019-04194.
		
		\bibliographystyle{elsarticle-harv}
		\bibliography{he}




	\end{document}